\documentclass[12pt,journal]{IEEEtran} 
\ifCLASSINFOpdf
\else
\fi

\usepackage{color}
\usepackage{amssymb}
\usepackage{amsthm}
\usepackage[cmex10]{amsmath}
\newcommand{\Eb}[1]{{ \mathbb{E}\left[ #1 \right] }}

\DeclareMathOperator{\C}{\mathbb{C}}

\newcommand {\Define} {\stackrel {\Delta} {=}  }

\newcommand{\mya}{\mathrel{\overset{\makebox[0pt]{{\tiny(a)}}}{=}}}

\newcommand{\myb}{\mathrel{\overset{\makebox[0pt]{{\tiny(b)}}}{=}}}

\newcommand {\pu} {p_{\text{u}}}

\newcommand{\N}{N_{\text{ZC}}}

\usepackage{bm}
\usepackage{mathtools}
\usepackage{fixltx2e}
\usepackage{epsfig,makeidx,color}
\usepackage{graphicx,dblfloatfix}
\usepackage{amsbsy}
\usepackage{amssymb}
\usepackage{euscript}
\usepackage{lipsum}
\usepackage{subfig}
\usepackage[ruled,linesnumbered,resetcount, norelsize]{algorithm2e}
\usepackage{cite}
\usepackage{nicefrac}
\usepackage{chngcntr}
\usepackage{lineno}
\usepackage[T1]{fontenc}
\usepackage{microtype}
\usepackage{wasysym}
\usepackage{footnote}
\usepackage[left=1 in,top=0.9 in,right= 1 in,bottom=0.9 in]{geometry}        

\newtheorem{proposition}{Proposition}

\newtheorem{remark}{\it Remark}

\usepackage[english]{babel}
  \usepackage[font=small,labelsep=space]{caption}
\renewcommand{\baselinestretch}{1.6}
\makeatletter
\def\citenoauxwrite#1{\begingroup
\@fileswfalse
\cite{#1}\relax
\endgroup}
\makeatother

\begin{document}

\title{Timing Advance Estimation and Beamforming of Random Access Response in Crowded TDD Massive MIMO Systems}
%
%
%

\author{Sudarshan~Mukherjee,~ 
Alok Kumar Sinha~and
        ~Saif Khan~Mohammed
\thanks{The authors are with the Department of Electrical Engineering, Indian Institute of Technology (I.I.T.) Delhi, India. Saif Khan Mohammed is also associated with Bharti School of Telecommunication Technology and Management (BSTTM), I.I.T. Delhi. Email: saifkmohammed@gmail.com. This work is supported by EMR funding from the Science and Engineering
Research Board (SERB), and also by the Visvesvaraya Young faculty Fellowship and PhD scheme of the Ministry of Electronics and Information Technology (MeitY) Govt. of India.}
}

\onecolumn
\maketitle

\vspace{-2.5 cm}

\begin{abstract}
Timing advance (TA) estimation at the base station (BS) and reliable decoding of random access response (RAR) at the users are the most important steps in the initial random access (RA) procedure. However, due to the limited availability of physical resources dedicated for RA, successful completion of RA requests would become increasingly difficult in high user density scenarios, due to contention among users requesting RA. In this paper, we propose to use the large antenna array at the massive MIMO BS to jointly group RA requests from different users using the same RA preamble. We then beamform the common RAR of each detected user group onto the same frequency resource, in such a way that most users in the group can reliably decode the RAR. The proposed RAR beamforming therefore automatically resolves the problem of collision between multiple RA requests on the same RA preamble, which reduces the RA latency significantly as compared to LTE.  Analysis and simulations also reveal that for a fixed desired SINR of the received RAR, both the required per-user RA preamble transmission power and the total RAR beamforming power can be decreased roughly by 1.5 dB with every doubling in the number of BS antennas.
\end{abstract}

\vspace{-1 cm}
\begin{IEEEkeywords}
\vspace{-0.5 cm}
Beamforming, random access response (RAR), massive MIMO, physical random access channel (PRACH), OFDM, timing advance, Zadoff-Chu (ZC) sequence.

\end{IEEEkeywords}

\vspace{-1.1 cm}

%
\IEEEpeerreviewmaketitle

\section{Introduction}
%
%
%
%

\vspace{-0.5 cm}

\par {In current communication systems (e.g. LTE) random access (RA) procedure is used by user equipments (UEs) to obtain dedicated physical resources for uplink (UL) communication. In the first step of the conventional RA procedure, each UE chooses a RA preamble at random and transmits it on a dedicated physical resource (e.g. PRACH (physical random access channel) in LTE) \cite{Baker, Skold}. If two or more UEs transmit the same RA preamble, then most likely the RA requests of all these UEs would collide, resulting in RA failure for almost all of these UEs. The number of distinct RA preambles is generally fixed and usually depends on the ratio of the time duration of the RA preamble to the maximum round-trip propagation delay in the cell \cite{Baker}.}

\par {In fifth generation (5G) communication systems, the connection density is expected to increase ten-fold as compared to 4G systems \cite{IMT2020}. As the number of RA preambles is generally fixed, this increase in connection density would increase chances of RA preamble collision, which would increase the number of repeat RA attempts, thereby increasing the average latency of the RA procedure in current communication systems (e.g. LTE). The UL timing of UEs requesting random access is not synchronized to the UL timing of the base station  (BS) and other synchronized UEs. Therefore in the second step of the conventional RA method, the BS estimates the UL timing/round-trip propagation delay for each UE which had transmitted some RA preamble in the first step, and whose RA preamble was successfully detected at the BS. Subsequently, for each detected RA preamble, the BS broadcasts the corresponding UL timing estimate and the scheduling grant information (also known as the random access response (RAR)) using dedicated downlink (DL) physical resource \cite{Baker, Skold}. Clearly, broadcasting of RAR in current communication systems is not as energy efficient as is required for 5G systems \cite{IMT2020}.}

\par {Massive multiple-input multiple-output (MIMO)/large scale antenna systems (LSAS) is one of the key 5G technologies, because of its characteristic ability to achieve very high energy and spectral efficiency\cite{Andrews1, Boccardi, Marzetta1, Marzetta2}. There are however few works which propose to use the large antenna array in massive MIMO (MaMi) systems to improve the energy efficiency of RA procedure and reduce  its latency (i.e. reduction in the number of repeat RA attempts). In \cite{Carvalho, Carvalho3}, a strongest user collision resolution (SUCR) mechanism exploiting the large antenna array at the BS is proposed for random pilot access in crowded mobile broadband (CMBB) scenarios. However, in SUCR, perfect UL timing synchronization is assumed and is therefore not applicable for initial random access.{\footnote[1]{{The differences between the SUCR and our proposed RA procedure have been clarified through footnotes~9 and~13.}}} In \cite{Lucas}, the authors propose a method for estimating the uplink timing in MaMi systems, which exploits the large antenna array at the BS to detect multiple RA requests from different UEs on the same RA preamble. However, this work assumes no time-frequency variation of the channel gains, which limits the available physical resource for RA preamble transmission, and hence this assumption limits the number of available distinct RA preambles. This will then result in more frequent collisions between RA preamble transmissions which will increase RA latency and decrease its energy efficiency (due to an increase in the number of repeat RA attempts).}

\par {In this paper, we propose a novel approach to the RA procedure for time division duplexed (TDD) MaMi systems, where we exploit the large antenna array at the BS to successfully detect multiple RA requests on the same RA preamble.\footnote[2]{{Unlike \cite{Lucas}, in our proposed RA procedure, the transmission of RA preamble spans multiple coherence bandwidths.}} Further, the channel reciprocity in TDD systems allows us to use the channel state information (CSI) acquired from the received RA preambles in the uplink, to simultaneously beamform the RAR from the BS to all UEs (detected on the same RA preamble), on the same frequency resource used for RA preamble transmission in the uplink. The proposed RA procedure can therefore successfully handle \emph{much higher connection densities} compared to the RA procedure in current communication systems, while maintaining \emph{a sufficiently low RA latency}. Beamforming of RAR using large antenna array helps in reducing the RAR transmit power significantly, while maintaining reliable detection of RAR at the UEs. In contrast to broadcasting of RAR, the proposed beamforming of RAR significantly improves the energy efficiency of the RA procedure.}

\par \textit{Contributions}: {The major contributions of our work are summarized in the following. \emph{Firstly}, in the proposed method, for each RA preamble, a time-correlation sequence between the received RA signals and the RA preamble is computed at each BS antenna (for UL timing estimation). In this paper, for each RA preamble we propose that \emph{the corresponding time-correlation sequences be averaged across the BS antennas} (spatial averaging). This reduces the effective noise power and \emph{allows for more than one UE to be detected on the same RA preamble} (see Section~\ref{taest}). Note that this scenario would have usually resulted in a collision in 4G systems. \emph{Secondly}, this reduction in the effective noise power further allows us to reduce the required per-user RA preamble transmission power, thereby improving the energy efficiency. For example, to achieve a fixed probability of UL timing estimation error, the minimum required per-user RA preamble transmit power can be decreased roughly by 1.5 dB with every doubling in the number of BS antennas (see Table~\ref{table:varMpu} in Section~\ref{taestimatenocont}, where we also see that with 80 BS antennas the required per-user RA preamble transmit power for the proposed RA method is about 30 dB less than that required by the LTE RA procedure). \emph{Thirdly}, for each RA preamble, we propose a \emph{novel grouping of UEs} that transmit the same RA preamble and have similar round-trip propagation delay between themselves and the BS. We propose to use the received RA preambles at the BS to estimate a common uplink timing and a common channel impulse response (CIR) for each such group of UEs (see Section~\ref{chanestsec}). Each group of UEs is then allocated a common time-frequency resource for subsequent UL transmission. The common UL timing estimate and the common scheduling information for each UE group is collectively called the \emph{group common RAR}. Our \emph{fourth contribution} is that, in Section~\ref{rarbeam} we propose to use the large antenna array at the BS to beamform the \emph{group common RAR} to the corresponding UE group. We show that in each such UE group, only those UEs would be able to reliably detect the RAR, whose CIR contribute significantly to the \emph{group common CIR} estimate. For instance, in a given group, UEs whose round-trip propagation delay differs from the group common timing estimate by more than the maximum channel delay spread, would not contribute significantly to the \emph{group common CIR} estimate and hence would most likely be unable to reliably decode the RAR. These UEs would then be automatically forced to re-initiate the RA procedure. This novel feature of the proposed RA method allows for \emph{automatic resolution of contention} among UEs transmitting the same RA preamble. We show that for a fixed UE density, our proposed RA procedure out-performs the LTE RA procedure both in terms of RA latency and energy efficiency. To be precise, with a fixed RA preamble transmit power and fixed RAR beamforming power, the average number of repeat RA attempts (equivalently the RA latency) of our proposed method is observed to decrease with increasing number of BS antennas. Analysis of the received SINR of RAR transmission at a UE reveals that, with every doubling in the number of BS antennas, both the per-user RA preamble transmission power and the total RAR beamforming power can be roughly decreased by 1.5 dB each, so that as the number of BS antennas asymptotically goes to infinity, the received SINR converges to a non-zero constant value, which does not depend on the UE density (see Propositions~\ref{PtvsM} and~\ref{finiteMvsKg} in Section~\ref{sinranalysis}). These results show the robustness of our proposed RA method in high UE density scenarios (e.g. CMBB etc.).} [\textbf{{Notations:}} $\C$ is the set of complex numbers, $\Eb{.}$ denotes the expectation operator. $(.)^{\ast}$ and $(.)^T$ denote conjugate and transpose operations respectively. $card(A)$ denotes the number of elements in set $A$.]

\vspace{-0.6 cm}

\section{Proposed Timing Advance Estimation}
\label{taest}
\vspace{-0.5 cm}
The round-trip propagation time delay between the base station and each UE is estimated at the BS. Since all UES are at different distances from the BS, the propagation time delay between the BS and each UE would  be different. This would cause unsynchronized reception of multi-user information signal at the BS in the uplink. The solution to this problem is to firstly estimate the round-trip propagation delay from each UE, and then feed this estimate back to the corresponding UEs. Based on the received estimate, each UE then advances its UL timing which ensures that in the subsequent UL slots, the uplink transmissions from all UEs are received at the BS in a time synchronized manner. As each UE \emph{advances} its UL timing based on the base station's estimate of the UE's round-trip propagation delay, this estimate is appropriately referred to as the \emph{timing advance} (TA). The TA estimation for a UE is performed based on the time of arrival of the RA preamble transmitted by that UE at the BS. In the following, in Section~\ref{preambleprocesstransmit} we first discuss the transmission of RA preambles from UEs and also the processing of the received RA preambles at each BS antenna. Next in Section~\ref{taestimatenocont}, we motivate the proposed spatial averaging based TA estimation algorithm, which is then presented in detail in Section~\ref{taestcont}.

\vspace{-0.8 cm}

\subsection{Preamble Sequence Transmission}\label{preambleprocesstransmit}
\vspace{-0.2 cm}

Each user intending to perform random access, chooses a RA preamble at random from a pre-determined set of preambles. As in LTE, in this paper also, we use RA preambles which are cyclically time-shifted versions of the basic root Zadoff-Chu (ZC) sequence. Subsequently, we denote the root ZC sequence by $s[t]$ ($t = 0, 1, \ldots, \N - 1$), where $\N$ is the length of this sequence. Let $K$ be the number of UEs requesting random access, with the $q^{\text{th}}$ UE ($q = 1, 2, \ldots, K$) transmitting a cyclically shifted version of $s[t]$, having cyclic shift $c_q \in [0, \N - 1]$, which we denote by $s_q[t]$, i.e.,

\vspace{-1.2 cm}

\begin{IEEEeqnarray}{rCl}
\label{eq:zcsequtq}
s_q[t] & \Define & s[(t - c_q) \hspace{-0.2 cm}\mod \hspace{-0.1 cm}\N] , \,\,\, 0 \leq t \leq \N-1 .
\IEEEeqnarraynumspace
\end{IEEEeqnarray}

\vspace{-0.4 cm}

\indent The ZC sequence is a constant envelope sequence ($|s[t]|^2 = 1$, $\forall t \in [0, \N -1]$), which has zero auto-correlation property, i.e., any two cyclically shifted versions of the same root ZC sequence having different shifts are orthogonal to each other\cite{Skold}. Clearly, from the property of the ZC sequence, we have

\vspace{-0.8 cm}

\begin{IEEEeqnarray}{rCl}
\label{eq:zccorr}
\sum\limits_{t=0}^{\N-1} s_q[t] \, s_k^{\ast}[t] & = & \left \{\begin{array}{ll}
\sum\limits_{t=0}^{\N-1}|s_q[t]|^2 = \N, & \text{if }c_q = c_k\\
0, & \text{otherwise}
\end{array} \right. \, .
\IEEEeqnarraynumspace
\end{IEEEeqnarray}

\vspace{-0.2 cm}

\indent Assuming the round-trip propagation delay of the $q^{\text{th}}$ UE to be $\tau_q$, the received ZC sequence transmitted by the $q^{\text{th}}$ UE when correlated with the root ZC sequence, would be detected in the correlation time-lag interval $[c_q + \tau_q, c_q + \tau_q + L -1]$, where $L$ is the channel delay spread (note that the cyclic shift of $c_q$ channel uses acts effectively as an extra propagation delay in addition to $\tau_q$). Similarly, for the $k^{\text{th}}$ UE with a round-trip propagation delay $\tau_k$ and using a cyclic shift $c_k$, the corresponding time-lag interval would be $[c_k + \tau_k, c_k +\tau_k + L-1]$. Note that if $c_q \neq c_k$ and the correlation time-lag intervals of these two users ($q^{\text{th}}$ and the $k^{\text{th}}$ UE) overlap, then it would be difficult to correctly estimate the timing advance for both these UEs. To avoid such situations, the users are allowed to choose the cyclic shift values only from a restricted set which is a subset of $\{0, 1, 2, \ldots, \N - 1\}$. Without loss of generality, let $c_q > c_k$, in which case, the above two time-lag intervals would not overlap if $(c_q - c_k) \mod \N \geq L + N_{roundtrip} \Define G$, where $N_{roundtrip}$ channel uses models the maximum round-trip propagation delay for any UE within the cell (i.e. $\tau_q \in [0, N_{roundtrip}= G - L]$).\footnote[3]{Subsequently, in the paper, we would denote the maximum round-trip delay by $G-L$.} The above condition $(c_q - c_k) \mod \N \geq G$ implies that the allowed cyclic shifts must be separated by at least $G$ channel uses, and therefore the number of allowed cyclic shifts is at most $Q \Define \lfloor \frac{\N}{G} \rfloor$.

\par Initially, there is no uplink timing synchronization and therefore the ZC sequence transmitted by each UE is followed by a guard time of duration at least $G$ channel uses. The last $G$ symbols of the ZC sequence  to be transmitted is also cyclic prefixed to the start of the RA preamble.\footnote[4]{This cyclic prefix of $G$ channel uses of the transmitted ZC sequence ensures that within the first ($G + \N$) channel uses from the start of the uplink slot for RA transmission, the complete ZC sequence is received at the BS for all the UEs.} The transmitted RA preamble from the $q^{\text{th}}$ UE, denoted by $x_q[t]$ ($t \in [0, \N + 2G - 1]$) thus consists of three parts, namely the cyclic prefix ($G$ channel uses), followed by the ZC sequence ($\N$ channel uses) and lastly the guard period ($G$ channel uses), i.e.,

\vspace{-1 cm}

\begin{IEEEeqnarray}{rCl}
\label{eq:txpreamble}
x_q[t] & = & \left\{ \begin{array}{ll}
s_{q}[t + N_{\text{ZC}} - G], & 0 \leq t \leq G-1 \hspace{0.5 cm} (\text{Cyclic Prefix})\\
s_{q}[t - G], & G \leq t \leq N_{\text{ZC}} + G - 1 \hspace{0.5 cm} (\text{ZC sequence})\\
0, & \N + G \leq t \leq \N + 2G - 1 \hspace{0.5 cm} (\text{Guard Period})
\end{array}\right. \,. 
\end{IEEEeqnarray}

\vspace{-0.2 cm}

\par The RA preamble thus received at the $m^{\text{th}}$ BS antenna from all $K$ UEs requesting random access is therefore given by

\vspace{-1.1 cm}

\begin{IEEEeqnarray}{rCl}
\label{eq:rxsigm}
y_m[t] & = & \sqrt{\pu}\sum\limits_{q=1}^{K}\sum\limits_{l=0}^{L-1} h_{mq}[l]x_q[t - l - \tau_q] \, + \, n_m[t] \, ,
\end{IEEEeqnarray}

\vspace{-0.2 cm}

\noindent where $t = 0, 1, \ldots, \N+2G-1$ and $\pu$ is the average per-user RA preamble transmission power. Note that $h_{mq}[l] \sim \mathcal{C}\mathcal{N}(0, \sigma_{hql}^2)$ models the $l^{\text{th}}$ tap of the channel impulse response (CIR) between the $m^{\text{th}}$ BS antenna and the $q^{\text{th}}$ UE. Further, note that $h_{mq}[l]$ ($l = 0, 1, \ldots, L-1$ and $m = 1, 2, \ldots, M$, where $M$ is the number of BS antennas) are modelled as statistically independent random variables. Here $\{\sigma_{hql}^2\}$, $l=0,1, \ldots,L-1$ denotes the power delay profile (PDP) for the $q^{\text{th}}$ UE. Finally, $n_m[t] \sim \mathcal{C}\mathcal{N}(0, \sigma^2)$ models the independent and identically distributed (i.i.d.) complex circular symmetric AWGN at the $m^{\text{th}}$ BS antenna ($m = 1,2, \ldots, M$).


\subsubsection*{{Processing of the Received RA Preamble at the $m^{\text{th}}$ BS Antenna}}

\vspace{-0.3 cm}

{The first $G$ samples received at the $m^{\text{th}}$ BS antenna from the start of the uplink RA slot (i.e. $y_m[t]$, $t \in [0, G-1]$) are ignored, since due to lack of timing synchronization, this part of the received signal might not contain RA preambles transmitted from all UEs requesting random access. Due to the addition of cyclic prefix of $G$ channel uses to the transmitted RA preamble (see \eqref{eq:txpreamble}), the next $\N$ samples (i.e. $y_m[t]$, $t = G, \ldots, G + \N - 1$) are guaranteed to contain the complete ZC sequence transmitted by all UEs. We denote these $\N$ samples by $r_m[t] = y_m[t + G]$, where $t = 0, 1, \ldots, \N-1$. Due to the addition of the cyclic prefix to the start of RA preamble (see \eqref{eq:txpreamble}), from \eqref{eq:rxsigm} we have}

\vspace{-0.95 cm}

{\begin{IEEEeqnarray}{rCl}
\label{eq:rxsigproc}
\nonumber r_m[t] & = & \sqrt{\pu}\sum\limits_{q=1}^{K}\sum\limits_{l=0}^{L-1}h_{mq}[l] x_q[t - l - \tau_q + G] + n_m[t+G]\\
\nonumber & \mya & \sqrt{\pu}\sum\limits_{q=1}^{K}\sum\limits_{l=0}^{L-1}h_{mq}[l] \, s_q[t - l - \tau_q] + n_m[t+G]\\
& \myb &  \sqrt{\pu}\sum\limits_{q=1}^{K}\sum\limits_{l=0}^{L-1}h_{mq}[l] \, s[(t - l - \tau_q - c_q) \mod \N] + n_m[t+G] \, ,
\IEEEeqnarraynumspace
\end{IEEEeqnarray}}

\vspace{-0.95 cm}

{\noindent where step (a) and step (b) follow from \eqref{eq:txpreamble} and \eqref{eq:zcsequtq} respectively. In order to estimate the propagation delay of the UEs, we first perform circular time-correlation of the received signal $r_m[t]$ with the root ZC sequence $s[t]$ at the $m^{\text{th}}$ BS antenna. This time-correlation sequence, denoted by $z_m[t]$ for the $m^{\text{th}}$ BS antenna is given by}

\vspace{-1 cm}

{\begin{IEEEeqnarray}{rCl}
\label{eq:tdcorr}
\nonumber z_m[t]  & \Define & \frac{1}{\sqrt{\N}} \, \sum\limits_{t^{\prime}=0}^{\N-1}r_m[t^{\prime}] s^{\ast}[(t^{\prime} - t) \mod \N]\\
\nonumber & \mya & \frac{\sqrt{\pu}}{\sqrt{\N}}\sum\limits_{q=1}^{K}\sum\limits_{l=0}^{L-1}h_{mq}[l] \sum\limits_{t^{\prime} = 0}^{\N-1}s[(t^{\prime} - l - \tau_q - c_q) \mod \N] s^{\ast}[(t^{\prime} - t) \mod \N]\\
\nonumber & & \hspace{2 cm} + \frac{1}{\sqrt{\N}}{\sum\limits_{t^{\prime}=0}^{\N-1} n_m[t^{\prime}+G] s^{\ast}[(t^{\prime} - t) \mod \N]} \\
& \myb & \sqrt{\N \, \pu}\sum\limits_{q=1}^{K} h_{mq}[t - c_q - \tau_q] \, + \, \underbrace{\frac{1}{\sqrt{\N}}\sum\limits_{t^{\prime}=0}^{\N-1} n_m[t^{\prime}+G] s^{\ast}[(t^{\prime} - t) \mod \N]}_{\Define \, w_m[t]} \, ,
\IEEEeqnarraynumspace
\end{IEEEeqnarray}}

\vspace{-0.65 cm}

{\noindent where step (a) and step (b) follow from \eqref{eq:rxsigproc} and \eqref{eq:zccorr} respectively. Note that $w_m[t] \sim \mathcal{C}\mathcal{N}(0, \sigma^2)$.} Since $h_{mq}[l]$ is non-zero only for $l = 0, 1, \ldots, L-1$, it follows that the term $h_{mq}[t - c_q - \tau_q]$ in \eqref{eq:tdcorr} is non-zero, only for $t \in [c_q + \tau_q, c_q+\tau_q +L-1]$, $\forall \, q = 1, 2, \ldots, K$. Further since $\tau_q \in [0, G-L]$, any non-zero contribution from the $q^{\text{th}}$ UE in $z_m[t]$ would only appear during time instances $c_q \leq t \leq c_q + G-1$. Subsequently, we refer to this time interval $[c_q, c_q + G - 1]$ of $G$ channel uses as the time-lag interval for the $q^{\text{th}}$ UE. By keeping the permissible cyclic shifts to be $G$ channel uses apart, it is ensured that the UEs using different RA preambles (i.e. having different cyclic shifts of the root ZC sequence) would contribute to samples of $z_m[t]$ lying in mutually exclusive time intervals. From \eqref{eq:tdcorr} it also follows that when the uplink per-user transmit power $\pu$ is sufficiently large compared to the noise floor, we can determine the time-lag interval of a UE from the time-correlation sequence in \eqref{eq:tdcorr} and the first time-lag value in the time-lag interval would give an estimate of the round-trip propagation delay for that UE. On the other hand, if the noise floor is high, the accuracy of TA estimation would degrade.

\vspace{-0.7 cm}

\subsection{Motivation for Spatial Averaging based Timing Advance Estimation}
\label{taestimatenocont}

In this section, we propose a novel TA estimation method for MaMi systems, whose objective is to improve timing estimation accuracy by exploiting the large antenna array at the BS. {In energy efficient 5G systems, the RA preamble transmission power is expected to be low, which will make it difficult to accurately estimate the TA, specially when it is based on the received RA preamble at a few BS antenna as in LTE. However, if we average the absolutely squared time-correlation sequence $z_m[t]$ computed at all $M$ BS antennas, then the effect of the independent noise terms across the antennas would average out resulting in a much reduced effective noise floor. At the same time, the proposed spatial averaging also leads to the hardening of the effective squared channel gains, which in turn increases the chances of the BS being able to detect the presence of the RA preamble transmissions.\footnote[5]{{If we have only one BS antenna, then it is quite possible that the channel gain between this single antenna of BS and a UE is poor, leading to undetected RA preamble transmission from that UE. On the other hand, if the BS has several antennas, then it is likely that there will be some BS antennas whose channel gain to this UE would be strong, thereby increasing the chances of detecting the UE's RA transmission.}}} In the following, we first present our proposed concept of spatial averaging based TA estimation in the contention-free scenario and later in Section~\ref{taestcont} we propose the complete TA estimation algorithm in detail for the contention scenario.


\par Let us assume that there is no contention among UEs attempting random access, i.e., each UE uses a different permissible cyclic shift to generate its RA ZC sequence. Let $\Xi \Define \{\xi_1, \xi_2, \cdots, \xi_{Q}\}$ be the set of permissible cyclic shifts to the root ZC sequence that can be used for RA ZC sequence generation and let $c_q \in \Xi$ denote the cyclic shift randomly chosen by the $q^{\text{th}}$ UE. The BS attempts to detect RA attempts made using only the permissible shifts of the root ZC sequence. For the $k^{\text{th}}$ permissible cyclic shift, it is clear that any RA  preamble transmission using the $k^{\text{th}}$ shift would contribute only to the interval $t \in [\xi_k, \xi_k + G -1]$ of the time correlation sequence $z_m[t]$ (see the discussion after \eqref{eq:tdcorr}). Let us consider the scenario where the $q^{\text{th}}$ UE attempts random access using the $k^{\text{th}}$ RA preamble, i.e., $c_q = \xi_k$. From \eqref{eq:tdcorr}, it therefore follows 

\vspace{-0.9 cm}

\begin{IEEEeqnarray}{rCl}
\label{eq:ktdcor}
 z_m[t + \xi_k] & = & \left\{ \hspace{-0.1 cm} \begin{array}{ll}
\sqrt{\N \, \pu} \, h_{mq}[t - \tau_q] \,+ \, w_m[t+\xi_k] , & t \in [\tau_q, \tau_q +L-1]\\
w_m[t + \xi_k], & t \in [0, \tau_q - 1]\,  \& \,\, t \in [\tau_q+L, G-1] \, .
\end{array}\right.
\IEEEeqnarraynumspace
\end{IEEEeqnarray}

\vspace{-0.2 cm}

\noindent Using \eqref{eq:ktdcor}, we now propose the spatially averaged absolutely squared time-correlation sequence for the $k^{\text{th}}$ cyclic shift ($\xi_k$), which is given by

\vspace{-0.9 cm}

\begin{IEEEeqnarray}{rCl}
\label{eq:tdmod}
V_k[t] & \Define & \Bigg (\frac{1}{M}\sum_{m=1}^{M}|z_m[t + \xi_k]|^2 \Bigg) - \sigma^2 =  \left\{\begin{array}{ll}
\N \, \pu \, \rho_{q,t} \, +  \, \omega_t + \, \eta_{t,q} \, , & \tau_q  \leq t \leq \tau_q + L-1\\
\omega_t , & \text{elsewhere}
\end{array}\right. \, ,
\IEEEeqnarraynumspace
\end{IEEEeqnarray}

\vspace{-0.25 cm}

\noindent where $t = 0,1, \ldots, G-1$; $\rho_{q,t} \Define \frac{1}{M}\sum\limits_{m=1}^{M}|h_{mq}[t-\tau_q]|^2$ and $\omega_t$ and $\eta_{t,q}$ are defined as below

\vspace{-0.86 cm}

\begin{IEEEeqnarray}{rCl}
\label{eq:rhoetaomega}
\eta_{t,q} & \Define & \frac{2\sqrt{\N \, \pu}}{M}\sum\limits_{m=1}^{M}\Re\{h_{mq}[l]w_m^{\ast}[t + \xi_k]\} \,\, \text{and}\,\,
\, \omega_t \, \Define \, \Bigg(\frac{1}{M}\sum\limits_{m=1}^{M}|w_m[t + \xi_k]|^2\Bigg) - \sigma^2 \, .
\IEEEeqnarraynumspace
\end{IEEEeqnarray}

\vspace{-0.16 cm}

\noindent Note that both the noise terms $\omega_t$ and $\eta_{t,q}$ are zero mean with variances as given below

\vspace{-1.1 cm}

\begin{IEEEeqnarray}{rCl}
\label{eq:varomegaeta}
\Eb{|\omega_t|^2} & = & \frac{\sigma^4}{M} \,\,\,\,\, \text{and}\,\,\,
\Eb{|\eta_{t,q}|^2} \, = \, \frac{2\N \sigma^2\pu \rho_{q,t}}{M} \, .
\IEEEeqnarraynumspace
\end{IEEEeqnarray}

\vspace{-0.3 cm}

\indent To estimate the round-trip propagation time delay $\tau_q$ of the $q^{\text{th}}$ UE (which is the only UE in this contention-free case discussed here), we exploit the fact that the contribution from the UE's RA transmission would lie in the time correlation interval $[\tau_q, \tau_q + L-1]$, i.e., the round-trip propagation delay is equal to the first time lag value of the time correlation interval in the signal $V_k[t]$ (see \eqref{eq:tdmod} and \eqref{eq:tdcorr}). To detect this time-correlation interval, we firstly propose to apply a threshold to $V_k[t]$ in order to eliminate the effect of noise, i.e.,

\vspace{-0.95 cm}

\begin{IEEEeqnarray}{rCl}
\label{eq:threshold}
P_k[t] & \Define & \left\{\begin{array}{ll}
V_k[t], & V_k[t] > \theta_0\\
0, & V_k[t] \leq \theta_0
\end{array}\right. \, ,
\IEEEeqnarraynumspace
\end{IEEEeqnarray}

\vspace{-0.2 cm}

\noindent where $\theta_0$ is an appropriate threshold. With an appropriately chosen value of $\theta_0$, from \eqref{eq:threshold}, it is clear that if there is no RA attempt using the $k^{\text{th}}$ RA preamble, then with high probability, $P_k[t] = 0$, $\forall \, t \in [0, G-1]$. Further, with an appropriately chosen threshold $\theta_0$, if the $q^{\text{th}}$ UE is the only UE transmitting the $k^{\text{th}}$ RA preamble, then from \eqref{eq:tdmod} and \eqref{eq:threshold}, we expect to have $P_k[t] > 0$, only for $t \in [\tau_q, \tau_q + L-1]$. Therefore a good timing advance estimate for this UE would be given by the location of the first non-zero value in $P_k[t]$, i.e.,

\vspace{-1.3 cm}

\begin{IEEEeqnarray}{rCl}
\label{eq:taest}
\widehat{\tau}_q & \Define & \min\limits_{{ t \in [0,\, G-1], \,\,P_k[t] > 0} } t \, .
\IEEEeqnarraynumspace
\end{IEEEeqnarray}

\vspace{-0.4 cm}

\subsubsection*{{Choice of Threshold $\theta_0$}}

{From the above discussions, it is observed that the accuracy of the above proposed RA attempt detection/TA estimation algorithm depends on the value of the threshold, $\theta_0$. From the definition of $V_k[t]$ and $P_k[t]$ in \eqref{eq:tdmod} and \eqref{eq:threshold} respectively, we note that a small value of $\theta_0$ could lead to detection of RA preambles, even when no preamble has actually been transmitted. This event is known as the \textit{false alarm scenario}. On the other hand, if the threshold $\theta_0$ is too high, then it is possible that no preamble transmission is detected, even when some UE has actually transmitted the said RA preamble. This event is commonly referred to as the \emph{missed detection scenario}. Clearly, we should choose a threshold such that both the false alarm probability ($P_F$) and the missed detection probability $(1 - P_D)$ are sufficiently small ($P_D$ is the detection probability). From \eqref{eq:tdmod}, we know that in the presence of $k^{\text{th}}$ RA preamble, $V_k[t]$ is equal to the sum of a term proportional to transmit power $\pu$ and other noise terms, whereas in the absence of any RA preamble transmission, $V_k[t]$ simply consists of the noise term $\omega_t$, $\forall t \in [0, G-1]$. From \eqref{eq:varomegaeta}, we know that the standard deviation of $\omega_t$ is $\frac{\sigma^2}{\sqrt{M}}$, i.e., with increasing $M$, the pdf (probability density function) of $\omega_t$ will become concentrated around its mean value of zero. This is due to the \emph{proposed spatial averaging} of $z_m[t+\xi_k]$ in \eqref{eq:tdmod}, due to which the effective noise $\omega_t$ is the average of $M$ i.i.d. random variables (see \eqref{eq:rhoetaomega}). This shows that if the threshold $\theta_0$ is kept constant, then with increasing $M$, $P_F$ would monotonically decrease as the standard deviation of $\omega_t$ decreases as $\frac{1}{\sqrt{M}}$. Therefore for a fixed desired $P_F$, with increasing $M$, we should be able to decrease the threshold $\theta_0$. This statement is made precise in the following proposition.}

\vspace{-0.5 cm}

{\begin{proposition}
\label{PfavsM}
\normalfont If $\theta_0 = \kappa \frac{\sigma^2}{\sqrt{M}}$, for some $\kappa > 1$, then $P_F \leq 1 - [1 - \frac{1}{\kappa^2}]^G$.
\end{proposition}}

\vspace{-0.5 cm}

{\begin{IEEEproof}
See Appendix~\ref{falsealarm}. \hfill \IEEEQEDhere
\end{IEEEproof}}

\noindent {From Proposition~\ref{PfavsM} it is clear that a fixed $P_F < 1 - [1 - \frac{1}{\kappa^2}]^G$ can be achieved even when the threshold is decreased as $\frac{1}{\sqrt{M}}$, with increasing $M$. Further, irrespective of $M$, any desirable $P_F$ can be achieved by suitably choosing a sufficiently large $\kappa > 1$.}

\vspace{-0.7 cm}

\begin{figure}[t]
\vspace{-0.35 in} 
\subfloat[][]{\includegraphics[width= 3.2 in, height= 1.9 in]{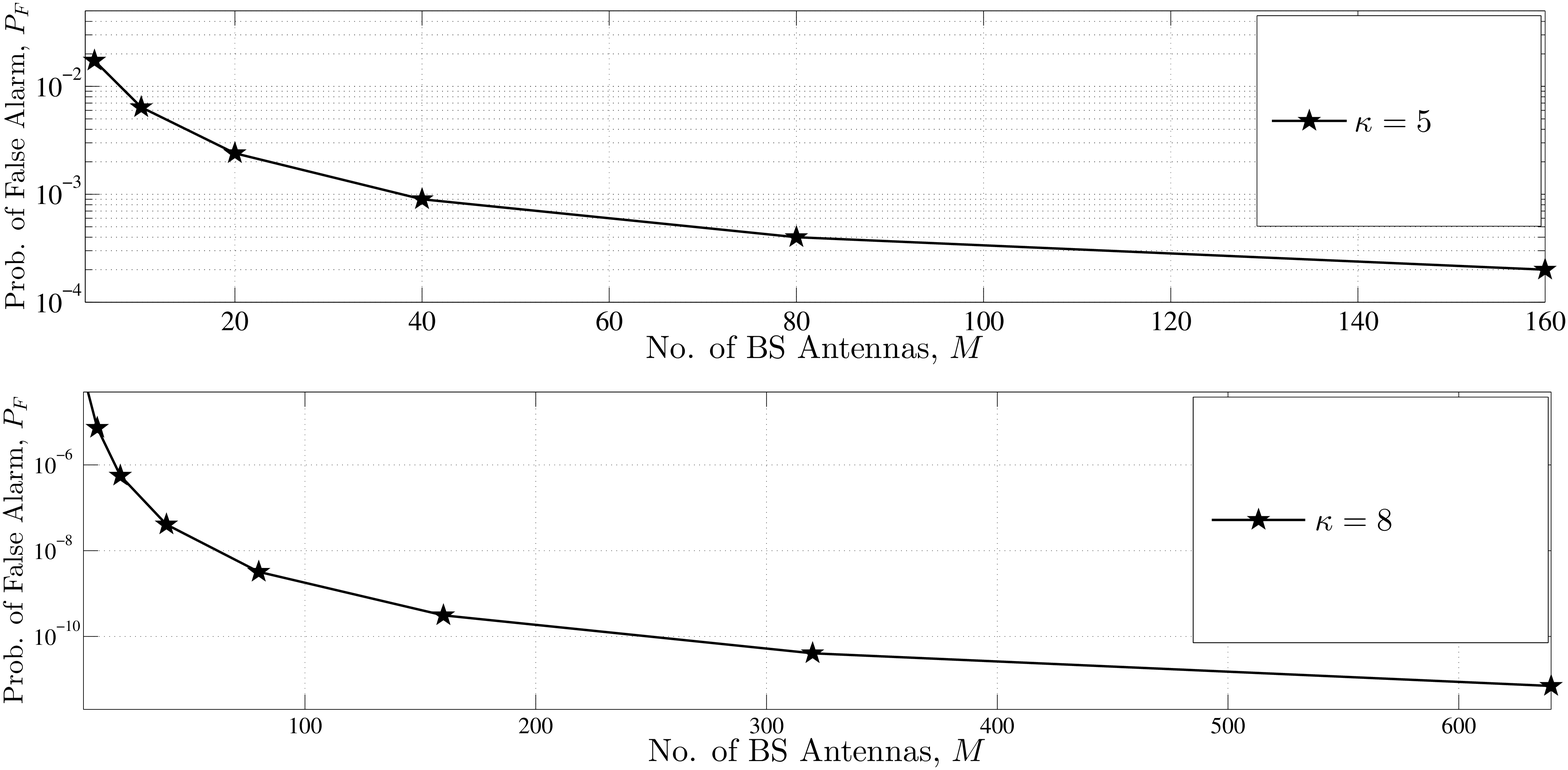}\label{fig:pfavsM1}}
\subfloat[][]{\includegraphics[width= 3.2 in, height= 1.9 in]{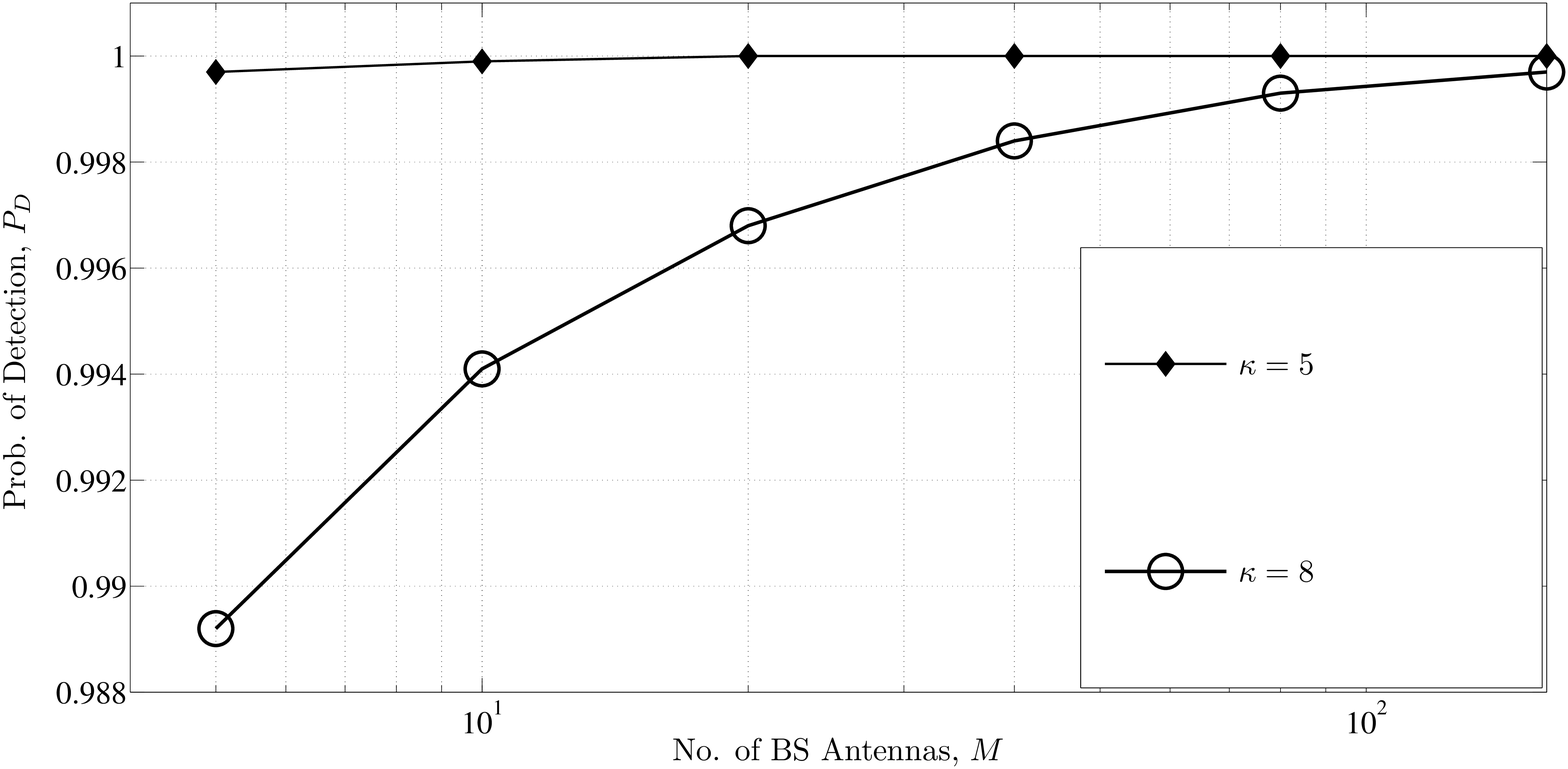}\label{fig:pfavsM2}}
\caption {{Plot of variation of (a) $P_F$ and (b) $P_D$ as a function of increasing number of BS antennas $M$, with $\frac{\theta_0}{\sigma^2} = \frac{\kappa}{\sqrt{M}}$ (fixed $\kappa = 5, 8$), $\N = 864$, and $\frac{\pu}{\sigma^2} = \frac{0.0632}{\sqrt{M}}$.}}
\label{fig:pfavsM}
\vspace{-0.6 cm}
\end{figure}

{\begin{remark}
\label{puthetaM}
\normalfont
{As MaMi systems are required to be energy efficient we would also like to decrease the RA preamble transmit power $\pu$ with increasing number of BS antennas, $M$. However, if $\pu$ decreases with increasing $M$, it is possible that the received RA preamble power (see the term $\N \pu \rho_{q,t}$ in \eqref{eq:tdmod}) would fall below the threshold $\theta_0$, leading to significant decrease in the detection probability $P_D$. Therefore with decreasing $\pu$, we must also reduce $\theta_0$ in order to maintain sufficiently high $P_D$. From Proposition~\ref{PfavsM} we know that for a fixed desired upper bound on $P_F$, $\theta_0$ can be decreased as $\frac{1}{\sqrt{M}}$, with increasing $M$. Therefore, it appears that we should also be able to decrease $\pu$ as $\frac{1}{\sqrt{M}}$ with increasing $M$, while maintaining a sufficiently high $P_D$ (see Fig.~\ref{fig:pfavsM}).} \hfill \qed
\end{remark}}
 
 \vspace{-0.3 cm}

 \begin{savenotes}
 \begin{table}[b]
 \caption[position=top]{{Min. reqd. $\frac{\pu}{\sigma^2}$ to achieve a fixed desired Prob. of TA estimation error $P_e = 10^{-2}$ and a fixed false alarm probability $P_F = 10^{-3}$, with increasing $M$, fixed $\N = 864$, $L = 6$ and $G = 50$ channel uses.}}
 \label{table:varMpu}
 \centering
 \begin{tabular}{| c | c | c | c | c| c|c|}
 \hline
 {$M$ (Number of BS antennas)} & {$20$} & {$40$} & {$80$} & {$160$} & {$320$} & {LTE ($M=1$)}\\ 
 \hline
 \vspace{-0.6 cm} & \\
 {Min. reqd. $\frac{\pu}{\sigma^2}$ (in dB)} & {-16.9} & {-19.35} & {-21.55} & {-23.5} & {-25.3}&{9}\\
 \hline
 \end{tabular}
 \end{table}
 \end{savenotes}

\par \textit{Impact of increasing number of BS antennas $M$ on the per-user UL Transmit Power, $\pu$}: The above discussion in Remark~\ref{puthetaM} has been supported through Table~\ref{table:varMpu}, where we numerically compute the minimum required $\pu$ for a fixed desired probability of false alarm and a fixed desired probability of timing estimation error (for the contention-free scenario, the timing estimate is said to be in error if and only if the actual value of the TA and its estimate are different).{\footnote[6]{{Note that the probability of TA estimation error is always greater than the probability of missed detection.}}} From Table~\ref{table:varMpu} it is observed that with $M \to \infty$, the required $\frac{\pu}{\sigma^2}$ decreases roughly by $1.5$ dB, with every doubling in the number of BS antennas $M$ (see the decrease from $M = 160$ to $M = 320$). {Here we also compute the minimum required $\frac{\pu}{\sigma^2}$ for the LTE TA estimation ($M = 1$). It is observed that the required $\frac{\pu}{\sigma^2}$ in LTE for the same desired performance is almost $30$ dB more than that required with $M = 80$ BS antennas by our proposed spatial averaging based TA estimation method for MaMi systems. Clearly, our proposed spatial averaging based TA estimation scheme is far superior to the conventional LTE TA estimation in terms of energy efficiency.}

\vspace{-0.8 cm}

\subsection{Timing Advance Estimation Algorithm}\label{taestcont}
\vspace{-0.3 cm}

In practice, the UEs requesting random access to the BS can randomly choose any one of the allowed/permissible RA preambles for transmission. Therefore it is possible that multiple UEs may use the same preamble for random access. Such scenario where multiple UEs use the same RA preamble is traditionally referred to as the \emph{contention scenario}. From \eqref{eq:tdcorr}, the time-domain correlation sequence computed at the $m^{\text{th}}$ BS antenna for the $k^{\text{th}}$ RA preamble is given by

\vspace{-1 cm}

\begin{IEEEeqnarray}{rCl}
\label{eq:ktdcorcont1}
z_m[t + \xi_k] & = & \sqrt{\N \, \pu}\sum\limits_{q = 1}^{K_k} h_{mq}[t - \tau_q] + w_m[t + \xi_k] \, ,
\end{IEEEeqnarray}

\begin{figure}[t]
\vspace{-0.6 cm}
\centering
\includegraphics[width= 6.4 in, height= 2.4 in]{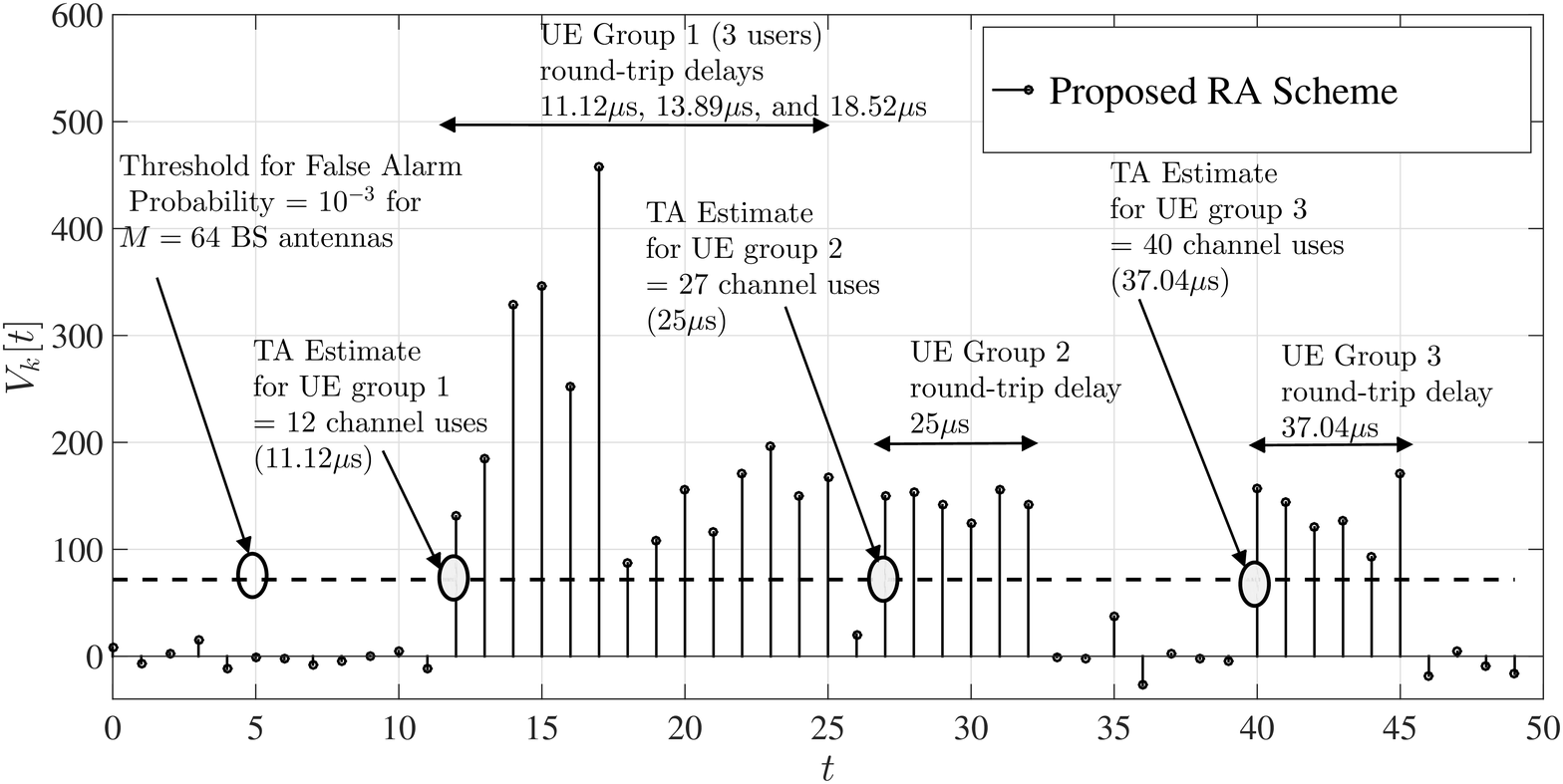}
\caption {{Contention scenario depicting user grouping and TA estimation for the $k^{\text{th}}$ RA preamble.}} 
\label{fig:uegroup1}
\vspace{-0.8 cm}
\end{figure}

\vspace{-0.2 cm}

\noindent where $t = 0, 1, \ldots, G-1$ and $K_k < K$ is the number of UEs transmitting the $k^{\text{th}}$ RA preamble. From $z_m[t + \xi_k]$ we then compute $V_k[t]$ and $P_k[t]$ as defined in \eqref{eq:tdmod} and \eqref{eq:threshold}. Clearly with an appropriately chosen threshold $\theta_0$ for a given $M$ (see the discussion on the choice of threshold in Section~\ref{taestimatenocont}), we have
$P_k[t]  =  \frac{\N \pu}{M}\sum_{m=1}^{M} \big|\sum_{q=1}^{K_k}h_{mq}[t - \tau_q]\big|^2 \, + \, \sum_{q=1}^{K_k}\eta_{t,q} \, + \, \omega_t$, when $V_k[t] > \theta_0$ and $P_k[t] = 0$ when $V_k[t] \leq \theta_0$. Here $\omega_t$ is defined in \eqref{eq:rhoetaomega} and $\eta_{t,q} = \frac{2\sqrt{\N \, \pu}}{M}\sum_{m=1}^{M}\Re\{h_{mq}[t- \tau_q]w_m^{\ast}[t+\xi_k]\}$. {From \eqref{eq:tdcorr} and \eqref{eq:ktdcor} it is clear that the time correlation between the received signal at the BS and the root ZC sequence would be non-zero at those time lags which fall within the $L$ length time correlation interval for some UE. As an example, in Fig.~\ref{fig:uegroup1}, we have plotted $V_k[t]$ versus $t$, where $5$ users (denoted as UE1, UE2, UE3, UE4 and UE5) transmit the $k^{\text{th}}$ RA preamble with $\frac{\pu}{\sigma^2} = -20.8$ dB, having individual round-trip delays $11.12 \mu$s, $13.89 \mu$s, $18.52 \mu$s, $25 \mu$s and $37.04 \mu$s respectively. Assuming a PRACH bandwidth of $1.08$ MHz as in LTE, the sampled round-trip delays would be $\tau_1 = 12, \tau_2 = 15, \tau_3= 20, \tau_4 = 27$ and $\tau_5 = 40$ channel uses. In Fig.~\ref{fig:uegroup1}, the threshold level $\theta_0$ is drawn with a dashed horizontal line. Clearly, with $L = 6$, the time correlation intervals for UE1, UE2, UE3, UE4 and UE5 are $[12, 17], [15, 20], [20, 25], [27, 32]$ and $[40, 45]$ respectively. It is therefore clear that with the appropriate choice of the threshold $\theta_0$, the non-zero samples of $P_k[t]$ would be in the time lag intervals $[12, 25]$, $[27, 32]$ and $[40, 45]$ respectively. If the time correlation intervals of UEs transmitting the same RA preamble are non-overlapping then the individual uplink timing information for all the UEs can be measured accurately. However, for the general case where the time correlation intervals of different UEs could overlap, we propose a novel user grouping based method for determining the timing information of all the UEs. We explain this method firstly through the example scenario in Fig.~\ref{fig:uegroup1} and then present it formally. Note that in Fig.~\ref{fig:uegroup1}, the time correlation intervals for UE1, UE2 and UE3 overlap with each other and hence they are grouped together as the first UE group. Similarly UE4 and UE5 form the second and third UE group respectively, since their time-correlation intervals are non-overlapping with each other and also with the correlation interval of the first UE group.}

\par In general, let $S$ UE groups be detected on the $k^{\text{th}}$ RA preamble, with the $g^{\text{th}}$ UE group consisting of $K_g$ UEs. Let the round-trip propagation delay for the $i^{\text{th}}$ UE in the $g^{\text{th}}$ UE group be denoted by $\tau_{g,i}$, and without loss of generality, we assume that $\tau_{g,1} \leq \tau_{g,2} \leq \cdots \leq \tau_{g, K_g}$. As the UEs in a group have overlapping time correlation intervals, it is clear that

\vspace{-1.2 cm}

\begin{IEEEeqnarray}{rCl}
\label{eq:Uegrp}
|\tau_{g, i} - \tau_{g, i+1}| \leq L, \, \forall i = 1, 2, \ldots, K_g - 1 \, .
\IEEEeqnarraynumspace
\end{IEEEeqnarray}

\vspace{-0.3 cm}

\indent From the above discussions, it is clear that the non-zero values in $P_k[t]$ due to transmission from UEs in the $g^{\text{th}}$ UE group would lie in the time correlation interval $[\tau_{g,1}, \tau_{g, K_g} + L-1]$. {Due to overlap of the time correlation intervals of UEs within a group, it is impossible to find the exact uplink timing (i.e. round-trip propagation delay) of all the UEs within the group. Hence, we propose that the starting time lag value of the time correlation interval of each UE group would be the uplink TA estimate for all the UEs in that UE group. This estimate would therefore be appropriately called the \emph{group common TA estimate} for that UE group. For instance, the location of the first non-zero sample in $P_k[t]$ would give the \textit{group common TA estimate} only for the \emph{first UE group} detected on the $k^{\text{th}}$ RA preamble. Note that this is essentially the estimate of the smallest round-trip delay in the first UE group and we denote it as $\widehat{\tau}_{1,1}$. In the scenario depicted in Fig.~\ref{fig:uegroup1}, we see that the group common TA estimate for the first UE group is $\widehat{\tau}_{1,1} = 12$ channel uses, which is also the estimate of the round-trip delay for UE1. From \eqref{eq:Uegrp} it is clear that non-zero samples of any other UE group in $P_k[t]$ can exist only after $t = \widehat{\tau}_{1,1}+L - 1$ (e.g. in Fig.~\ref{fig:uegroup1}, the time-correlation interval for the second UE group begins from $t = 27$-th channel use ($> \widehat{\tau}_{1,1} + L-1 = 17$)). Therefore, to mark the end of the time correlation interval  for the first UE group, we need to find the location of the first zero sample in $P_k[t]$, for $t > \widehat{\tau}_{1,1}+L-1$. Upon detection of this zero sample, we can re-initiate our search for the next UE group in the remaining part of $P_k[t]$, in a similar fashion as we did before for the first UE group. For instance, in Fig.~\ref{fig:uegroup1}, the group common TA estimates for the second and third UE groups are given by $\widehat{\tau}_{2,1} = 27$ and $\widehat{\tau}_{3,1} = 40$ respectively. Note that the successful detection of multiple UE groups on the same RA preamble is possible only because of the reduction in the effective noise, which is in turn due to the proposed spatial averaging of the squared time correlation sequences computed at each BS antenna. The proposed UE grouping method described above has also been summarized in Algorithm~\ref{alg:algo2}. It is also noted that the proposed UE grouping and the group common TA estimates are novel and are feasible only due to the fact that we exploit the large antenna array at the BS for the proposed spatial averaging.}


\begin{algorithm}[t]
\caption{Proposed UE grouping and TA estimation for the $k^{\text{th}}$ RA preamble.} 
\label{alg:algo2}
\textsc{Input}: $V_k[t] = \left(\frac{1}{M}\sum\limits_{m=1}^{M}|z_m[t+\xi_k]|^2\right) - \sigma^2$; 
\par  \textsc{Output}: \text{NumGrp}, $\widehat{\tau}_{g,1}$, $g = 1, \ldots, \text{NumGrp}$.

\par \textsc{STEP-1}: $P_k[t] = V_k[t]$, $t = 0, 1, \ldots, G-1$.

\par \textsc{Step-2}: for $t = 0:1:G-1$\\
						\hspace{2 cm}			 
						if  ($P_k[t] \leq \theta_0$)
						\hspace{2.5 cm}   
						   $P_k[t] = 0$;\\ 
						\hspace{2 cm}	     end \hspace{2.5 cm} \%End of if statement\\
						\hspace{1.5 cm} end \hspace{3 cm} \% End of for-loop
\par \textsc{Step-3}: Initialize $t = 0$, $g = 0$.

\par \textsc{Step-4}: While $t <= G - L$\\
					\textit{NzeroChk1}:	\hspace{1 cm}              if ($P_k[t] = 0$)
						\hspace{3.2 cm}                   $t = t + 1$;\\
						\hspace{3.2 cm}                 else \% UE group detected\\
						\hspace{4.4 cm}
						$g = g + 1$;\\
						\hspace{4.4 cm}
						$\widehat{\tau}_{g,1} = t$;
						\hspace{0.5 cm}
						$t = t + L$;\\
					\textit{ZeroChk2}:	\hspace{2.5 cm}
						While ($P_k[t] > 0$) \& ($t  <= G - L$)\\
						\hspace{6 cm}
						$t = t + 1$;\\
						\hspace{4.5 cm}
						end \hspace{2 cm}\% End of inner While loop starting at \emph{ZeroChk2}\\
						\hspace{3.5 cm}
						end \hspace{3 cm}\% End of \textit{NzeroChk1}\\
						\hspace{2 cm}  
						end \hspace{4 cm}\% End of outer While loop starting at STEP-4
\par \textsc{STEP-5}:  NumGrp = $g$;
\end{algorithm}

\vspace{-0.1 cm}

\subsubsection*{Complexity of the TA Estimation Algorithm} Note that the above proposed TA estimation algorithm first computes the time-domain (TD) correlation sequence of the received RA preambles at each BS antenna (see \eqref{eq:tdcorr}). The total number of complex operations required to compute this TD correlation sequence at each BS antenna is $\mathcal{O}(\N)$ (since the length of the preamble sequence is $\N$). Next, these TD correlations computed at each BS antenna are absolutely squared and then averaged across all $M$ antennas (see \eqref{eq:tdmod}). Thus the total number of complex operations required to compute $V_k[t]$ in \eqref{eq:tdmod} is $\mathcal{O}(M\N)$. Next the proposed user grouping  and TA estimation for UEs attempting RA using the $k^{\text{th}}$ RA preamble requires search on $\{P_k[t]\}$, a sequence of length $G$ channel uses (see lines 9-18 in STEP-4 of Algorithm~\ref{alg:algo2}). Since there are $Q = \lfloor \frac{\N}{G}\rfloor$ RA preambles, the total number of operations required to search for all $Q$ preambles is $\mathcal{O}(G\, Q) = \mathcal{O}(\N)$. Therefore the total number of operations required for TA estimation and the proposed user grouping is $\mathcal{O}(M \N + \N) = \mathcal{O}(M\N)$. Since the time correlation sequence is $\N$ channel uses long, the per-channel use complexity would be $\mathcal{O}(M)$ only, i.e., the complexity of the proposed user grouping and TA estimation algorithm increases only linearly with $M$.

\vspace{-0.7 cm}

\section{Downlink Beamforming for RAR Transmission}
\vspace{-0.35 cm}

After TA estimation, the BS is required to transmit the random access response (i.e. TA estimate, scheduling grant information etc.) to the UEs requesting random access for timing correction and subsequent uplink transmission. Since the UEs do not have any way of identifying whether their random access has been successful or not, they wait for the RAR from the BS in the downlink. {Conventionally, in LTE systems, once a random access attempt on a given RA preamble is detected, the BS first estimates the corresponding TA information. After that, it transmits the RAR for the detected RA preambles over the physical downlink shared channel (PDSCH) by using transmit diversity (e.g. SFBC/FSTD \cite{Skold}). For each detected RA preamble, the location of the PDSCH sub-carriers is however transmitted over the physical downlink control channel (PDCCH) along with the identifier of that RA preamble. From the received RA preamble identifier, the UE identifies the location of its corresponding RAR and upon successful RAR decoding, it uses the received TA estimate for timing correction.} Note that this \emph{RA preamble identifier-based two step approach of LTE RA procedure} by which a UE is able to identify its RAR would not work in \emph{our proposed user grouping and TA estimation algorithm based RA procedure}, since each RA preamble detected at the BS may have multiple UE groups with each UE group having a different RAR due to different group common TA estimates.

\par As more number of permissible RA preambles are likely to be transmitted in crowded scenarios, the two-step approach of the LTE RA procedure would also increase latency due to the limited availability of PDCCH resource.{\footnote[7]{{In LTE systems, each RA preamble can detect at most one UE and therefore for each RA preamble the BS broadcasts a single RAR. Due to limited PDCCH resource, the LTE RA procedure will be unable to handle the large number of RA requests in crowded scenarios. Also, since at most one UE can be detected on a RA preamble, the other UEs will be forced to repeat the random access requests by transmitting a randomly chosen RA preamble on the next available PRACH. With a large number of simultaneous RA requests, it is therefore clear that many UEs might have to repeat RA attempts which would increase the RA latency and also degrade the energy efficiency.}} With the proposed user grouping and TA estimation method, this problem of limited downlink resource for RAR transmission is even more problematic due to the possibility of many different RAR messages since each UE group has a different RAR response. Hence to address this issue of minimizing the amount of downlink physical resource required for the transmission of the RAR, in this paper, we propose to jointly beamform the RARs for all UE groups detected on a RA preamble, onto a dedicated downlink frequency resource, which is part of the frequency resource used by the PRACH in the uplink. Beamforming of RAR however requires the knowledge of channel state information (CSI) at the BS. We propose to use the received RA preambles in the uplink slot to estimate the CSI for each detected UE group. This CSI estimate is then used for joint beamforming of RAR.\footnote[8]{Estimating CSI from UL RA preambles is possible due to the channel reciprocity in TDD systems.} Thus, by sending the RAR over the same frequency resources as used by PRACH, we avoid burdening the PDCCH resource. Further the proposed downlink beamforming of RAR using the large antenna array in TDD MaMi systems gives high power gain which is not possible in LTE due to the \emph{lack of CSI and presence of only a few antennas at the BS}. Since the proposed RAR beamforming allows transmission of RAR for several UEs simultaneously, it also reduces the overall latency of the RA procedure and enables handling of a much larger number of simultaneous RA requests as compared to LTE.}

\vspace{-0.8 cm}

\subsection{Channel Estimation for UE Groups detected on the $k^{\text{th}}$ RA Preamble}
\label{chanestsec}
\vspace{-0.3 cm}

From \eqref{eq:ktdcorcont1}, it is clear that we can acquire an estimate of the channel impulse response (CIR) for individual UE groups detected on the $k^{\text{th}}$ RA preamble from $z_m[t+\xi_k]$ ($m = 1,2, \ldots, M$; $t = 0,1, \ldots, G-1$; and $k = 1, 2, \ldots, Q$). From our previous discussions on the proposed user grouping and TA estimation algorithm in Section~\ref{taestcont}, we note that in $P_k[t]$, the non-zero time correlation lag values of all UEs in the $g^{\text{th}}$ UE group detected on the $k^{\text{th}}$ RA preamble overlap with each other and are limited to the interval $\tau_{g,1} \leq t \leq \tau_{g, K_g}+L-1$, where $K_g$ is the number of UEs in the $g^{\text{th}}$ UE group (see the discussion after \eqref{eq:Uegrp}). Overlapping time-correlation intervals of UEs in a UE group implies that their CIRs would also overlap in time. Hence from the computed time correlation sequence $\{z_m[t+\xi_k]\}$ (starting at $t = \widehat{\tau}_{g,1}$), we propose to estimate a single combined channel impulse response for the entire group, which we subsequently refer to as the \emph{group common CIR} for all UEs in the $g^{\text{th}}$ UE group detected on the $k^{\text{th}}$ RA preamble. {Towards estimating the group common CIR for the $g^{\text{th}}$ UE group, we propose to use only the first $L$ samples, i.e., the samples of $z_m[t + \xi_k]$ for $t \in [\widehat{\tau}_{g,1}, \widehat{\tau}_{g,1} + L -1]$, as it ensures that the length of the estimated CIR is not more than the channel delay spread ($L$). If we allow more than $L$ samples to be used for CIR estimation, then it is possible that the RAR could be successfully decoded at some UEs having a round-trip delay which is $L$ channel uses more than the group common TA estimate, since our proposed group common TA estimate is the first time lag value of the time correlation interval of the UE group. For such a UE, RAR decoding and subsequent uplink timing correction would still result in an uplink timing error greater than $L$, which is the length of the cyclic prefix (CP) used in uplink OFDM transmission. This would then adversely affect the orthogonality between the sub-carriers leading to inter-carrier interference. Rewriting \eqref{eq:ktdcorcont1} in terms of the UE groups detected on the $k^{\text{th}}$ RA preamble, we have}

\vspace{-1 cm}

{\begin{IEEEeqnarray}{rCl}
\label{eq:ktdcorcont11}
\nonumber z_m[t + \xi_k] & = & \sqrt{\N \, \pu}\sum\limits_{q=1}^{K_k}h_{mq}[t - \tau_q] + w_m[t + \xi_k]\\
& = &  \sqrt{\N \, \pu}\sum\limits_{g=1}^{D_k}\sum\limits_{i=1}^{K_g}h_{mgi}[t - \tau_{g,i}] + w_m[t + \xi_k] \, ,
\IEEEeqnarraynumspace
\end{IEEEeqnarray}}

\vspace{-0.9 cm}

\noindent {where $D_k$ is the number of UE groups detected on the $k^{\text{th}}$ RA preamble and $h_{mgi}[l] \sim \mathcal{C}\mathcal{N}(0, \sigma_{hgil}^2)$ ($l = 0,1, \ldots, L-1$) is the complex baseband CIR between the $m^{\text{th}}$ BS antenna and the $i^{\text{th}}$ UE of the $g^{\text{th}}$ UE group. Clearly, the least square (LS) estimate of the group common CIR for the $g^{\text{th}}$ UE group detected on the $k^{\text{th}}$ RA preamble is computed as follows}

\vspace{-1.1 cm}

\begin{IEEEeqnarray}{rCl}
\label{eq:chanest1}
\widehat{h}_{m,g}[l] & = & \frac{1}{\sqrt{\N}} z_m[\widehat{\tau}_{g,1}+l + \xi_k] \, ,
\IEEEeqnarraynumspace
\end{IEEEeqnarray}

\vspace{-0.25 cm}

\noindent where $l = 0, 1, \ldots, L-1$. Substituting $z_m[t+\xi_k]$ from \eqref{eq:ktdcorcont11} in \eqref{eq:chanest1}, we get

\vspace{-0.9 cm}

\begin{IEEEeqnarray}{rCl}
\label{eq:chanest}
\widehat{h}_{m,g}[l] & = & \sqrt{\pu}\sum_{i=1}^{K_g}h_{mgi}[\Delta \tau_{g,i} + l] \, + \, \frac{1}{\sqrt{\N}}w_m[\widehat{\tau}_{g,1} + l + \xi_k]
\IEEEeqnarraynumspace
\end{IEEEeqnarray}

\vspace{-0.35 cm}

\noindent where $\Delta \tau_{g,i} \Define \widehat{\tau}_{g,1} - \tau_{g,i}$ is the timing error for the $i^{\text{th}}$ UE in the $g^{\text{th}}$ UE group. {For each UE group, its group common CIR is estimated only from the first $L$ samples of the corresponding time correlation interval of that group and hence it is clear that there will be some UEs in that UE group whose CIR will contribute partially to this group common CIR estimate and there will also be some UEs in that group, whose CIR will not at all contribute to the group common CIR estimate.\footnote[9]{{Our proposed RA method differs from the SUCR protocol for random pilot access in \cite{Carvalho}, as in \cite{Carvalho} it is assumed that the uplink transmission from all UEs is already perfectly synchronized, due to which complete CSI is obtained for all users. This is however not true for the initial access problem considered by us in this paper.}} We explain this briefly with the help of the example in Fig.~\ref{fig:uegroup1}.} {In Fig.~\ref{fig:uegroup1}, we see that three UE groups have been detected on the $k^{\text{th}}$ RA preamble, where the $1^{\text{st}}$ UE group has three UEs (round-trip delays 12, 15 and 20 channel uses) with a group common TA estimate of $t = 12$ channel uses. For a maximum delay spread of $5\mu$s and PRACH bandwidth $1.08$ MHz, the channel delay spread is $L = \lceil 1.08 \times 5 \rceil = 6$ channel uses. Clearly, the \emph{group common CIR} estimate for the $1^{\text{st}}$ UE group would be obtained from the samples of the correlation sequence $z_m[t + \xi_k]$ in the time interval $t \in [12, 17]$. Similarly, the \emph{group common CIR} estimate for the $2^{\text{nd}}$ and $3^{\text{rd}}$ UE groups would be obtained from the time intervals $[27, 32]$ and $[40, 45]$ respectively. Note that in the first UE group the group common CIR estimate is derived from the time-correlation sequence in the time lag interval $[12, 17]$ and since the time-lag interval corresponding to the RA preamble received from UE1 is also $[12, 17]$, the group common CIR estimate would contain the complete CIR of UE1 (see Fig.~\ref{fig:uegroup1}). For UE2 the time lag interval corresponding to its received RA preamble is $[15, 20]$ and therefore the group common CIR estimate would contain only that part of the CIR of UE2 which is in the time-lag interval $[15, 17]$. Finally, the time-lag interval corresponding to the RA preamble received from UE3 is $[20, 25]$ and therefore the CIR of UE3 would not at all contribute to the group common CIR estimate.}

\vspace{-0.75 cm}

\subsection{RAR Transmission: Frequency Domain Beamforming}
\label{rarbeam}
\vspace{-0.3 cm}

Once the channel estimates are acquired from the received RA preambles, the BS can beamform the RAR for all detected UE groups over the same frequency resource used by PRACH. The RAR for any of the detected UE groups would contain at least the following information: (a) random access (RA) acknowledgement; (b) \emph{group common TA estimate} for that UE group; and (c) resource allocation/scheduling grant (i.e. location of allocated subcarriers for subsequent UL transmission). Due to the small size of the RAR block, RAR transmission for any UE group does not require the entire PRACH bandwidth. Therefore, in order to reduce multi-user interference (MUI), we can schedule RAR transmission for UE groups detected on different RA preambles onto different subcarriers. {For instance, let us assume that the PRACH has $N_{\text{RS}}$ shared channel (SCH) subcarriers and for each RA preamble detected at the BS, a dedicated portion (say $N_{\text{SC}}$ subcarriers) of this overall bandwidth is allocated for the downlink beamforming of RAR. Also, for each RA preamble, the RARs of different UE groups detected on this RA preamble are simultaneously beamformed on the same frequency resource. Assuming $N_{\text{SC}}$ subcarriers to be sufficient for complete transmission of the RAR sequence of any UE group, the minimum number of OFDM symbols required for RAR transmission for all $Q$ RA preambles would be $N_{\text{slot}} = \lceil \frac{N_{\text{SC}} \, Q}{N_{\text{RS}}} \rceil$.\footnote[10]{{Here we assume that for a given RAR sequence of a UE group, each allocated subcarrier carries only one RAR symbol, i.e., with $N_{\text{SC}}$ allocated subcarriers, the maximum length of the RAR sequence would also be $N_{\text{SC}}$.}}}

{ Next we discuss the proposed RAR beamforming for the $g^{\text{th}}$ UE group detected on the $k^{\text{th}}$ RA preamble. Here we assume that there are $D_k$ UE groups detected on the $k^{\text{th}}$ RA preamble and $S_k$ is the set of indices of subcarriers allocated for RAR transmission to UE groups detected on the $k^{\text{th}}$ RA preamble, i.e., $card(S_k) = N_{\text{SC}}$. Let $u_g[n]$ be a symbol of the group common RAR of the $g^{\text{th}}$ UE group (detected on the $k^{\text{th}}$ RA preamble) which will be transmitted on the $n^{\text{th}}$ subcarrier ($n \in S_k$). We propose to use conjugate beamforming/maximum ratio transmission (MRT) to precode $u_g[n]$ onto the signal to be transmitted from each BS antenna. For the $g^{\text{th}}$ UE group the signal transmitted from the $m^{\text{th}}$ BS antenna on the $n^{\text{th}}$ subcarrier is then given by\footnote[11]{{We do not use a subscript $k$ in the notation for this transmit signal for the sake of simplicity.}}}

\vspace{-1.3 cm}

{\begin{IEEEeqnarray}{rCl}
\label{eq:transmitsignal}
X_{m,g}[n] & \Define & \frac{1}{\sqrt{\upsilon_g}}\widetilde{H}_{m,g}^{\ast}[n] \, u_g[n] \, ,
\end{IEEEeqnarray}}

\vspace{-1 cm}

\noindent {where $\upsilon_g \Define {\Eb{||\widetilde{\bm H}_{g}[n]||^2}}$ and $\widetilde{\bm H}_g[n] \Define (\widetilde{H}_{1,g}[n], \cdots, \widetilde{H}_{M,g}[n])^T$. Here $\widetilde{H}_{m,g}[n]$ is our proposed estimate of the frequency domain channel gain on the $n^{\text{th}}$ subcarrier which is given by}

\vspace{-1.2 cm}

\begin{IEEEeqnarray}{rCl}
\label{eq:fdchanest}
\widetilde{H}_{m,g}[n] & = & \frac{1}{\sqrt{N_{\text{RS}}}}\sum\limits_{l=0}^{L-1} \widehat{h}_{m,g}[l]e^{-j\frac{2\pi}{N_{\text{RS}}}nl} \, ,
\IEEEeqnarraynumspace
\end{IEEEeqnarray}

\vspace{-0.35 cm}

\noindent where $\widehat{h}_{m,g}[l]$ is defined in \eqref{eq:chanest}. Note that this estimate is derived from the $N_{\text{RS}}$-point DFT of the estimated time-domain group common CIR of the $g^{\text{th}}$ UE group (see Section~\ref{chanestsec}).\footnote[12]{We assume that the uplink slot used for transmission of the RA preamble and the DL slot used for RAR transmission lie in the same coherence interval.} {As there are $D_k$ UE groups on the $k^{\text{th}}$ RA preamble and since $n \in S_k$, the total signal transmitted by the $m^{\text{th}}$ BS antenna on the $n^{\text{th}}$ subcarrier is given by\footnote[13]{{In the proposed RAR beamforming method, different RARs for different UE groups detected on the same RA preamble are jointly beamformed on the same time-frequency resource. This is however different from the SUCR protocol in \cite{Carvalho} where the same signal is sent to all users who used the same pilot during the UL slot.}} $X_{m}[n] = \sqrt{P_{\text{d}}}\sum\limits_{g=1}^{D_k}X_{m,g}[n]$, where $P_{\text{d}} \Define P_T\frac{N_{\text{RS}}}{N_{\text{SC}}K_t}$. Here $K_t \Define \sum\limits_{k=1}^{Q}D_k$ is the total number of UE groups detected on all $Q$ RA preambles and $P_T$ is the total downlink power transmitted by the BS. Finally at the $m^{\text{th}}$ BS antenna, $N_{\text{RS}}$-point IDFT of the frequency domain signal $X_m[n]$ ($n = 0,1, \ldots, N_{\text{RS}} -1$) is performed followed by addition of a $L$-length cyclic prefix before transmission. Note that the RAR symbols $u_g[n]$ are assumed to be of unit energy, i.e., $\Eb{|u_g[n]|^2} = 1$.}{\footnote[14]{{Since all UEs belonging to a detected UE group have the same common TA estimate and are scheduled on the same uplink resource, their RAR would not require any user dependent information.}}} Finally the signal received on the $n^{\text{th}}$ subcarrier, at the $i^{\text{th}}$ UE of the $g^{\text{th}}$ UE group is given by (after removal of CP and taking $N_{\text{RS}}$-point DFT)\footnote[15]{{Note that both the BS and the UEs are aware of the association/mapping between a permissible RA preamble and the set of downlink subcarriers allocated for the transmission of RAR to UE groups detected on this preamble. As each UE knows the RA preamble transmitted by it, it is aware of the subcarriers on which it should expect the RAR from the BS.}}

\vspace{-0.9 cm}

\begin{IEEEeqnarray}{lCl}
\label{eq:rxdcig}
Y_{g,i}[n]  & = &  \sum\limits_{q=1}^{D_k} \sqrt{\frac{N_{\text{RS}} \, P_{\text{d}}}{\upsilon_q}} \bm H_{gi}^T[n]\widetilde{\bm H}_{q}^{\ast}[n] u_q[n] \,+\, E_{g,i}[n] \,,
\IEEEeqnarraynumspace
\end{IEEEeqnarray}

\begin{figure}[t]
\vspace{-0.5 cm}
\centering
\includegraphics[width= 4.4 in, height= 2 in]{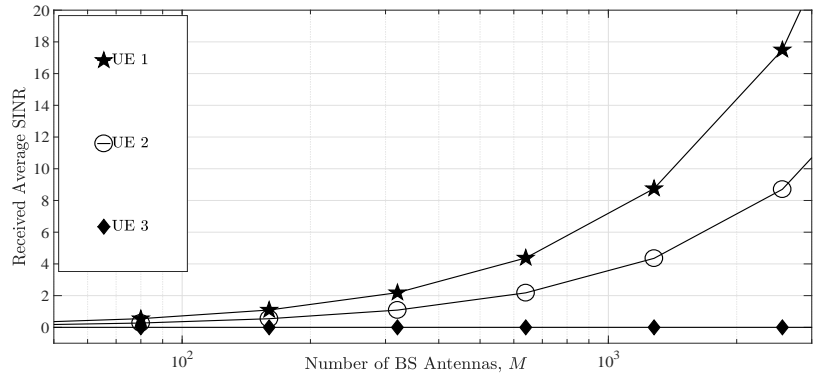}
\caption {{Plot of the average received SINR for all three UEs in UE group 1, for the scenario illustrated in Fig.~\ref{fig:uegroup1}.}} 
\label{fig:example}
\vspace{-0.4 cm}
\end{figure}

\vspace{-0.3 cm}

\noindent where $E_{g,i}[n] \sim \mathcal{C}\mathcal{N}(0, \sigma^2)$ is the complex circular symmetric baseband AWGN noise and $\bm H_{gi}[n] \Define (H_{1gi}[n], \cdots, H_{Mgi}[n])^T$. Here $H_{mgi}[n] \Define \frac{1}{\sqrt{N_{\text{RS}}}} \sum_{l=0}^{L-1}h_{mgi}[l]e^{-j\frac{2\pi}{N_{\text{RS}}}nl}$ is the frequency domain channel gain of the $n^{\text{th}}$ subcarrier between the $m^{\text{th}}$ BS antenna and the $i^{\text{th}}$ UE of the $g^{\text{th}}$ UE group detected on the $k^{\text{th}}$ RA preamble ($h_{mgi}[l]$ is defined in the line following \eqref{eq:chanest}). {We have earlier seen that the group common CIR estimate contains partial/incomplete CIR for some UEs in that UE group whose received signal-to-interference-and-noise ratio (SINR) will clearly get impacted. To study this, in Fig.~\ref{fig:example}, we plot the average received SINR for all three UEs in the first UE group for the example scenario illustrated in Fig.~\ref{fig:uegroup1}, as a function of increasing number of BS antennas, $M$ and fixed $\frac{\pu}{\sigma^2} = \frac{P_T}{\sigma^2} = -20.8$ dB. It is observed that for any given $M$, UE1 has the highest average received SINR, followed by UE2 and then UE3. This is so because the group common CIR estimate contains the complete CIR for UE1, while only partial CIR is present for UE2 and therefore the received SINR at UE2 is expected to be smaller than the received SINR at UE1. The received SINR for UE3 is even smaller than that of UE2, since there is no contribution of its CIR to the group common CIR estimate. Therefore, in this example scenario, it is clear that only UE1 and UE2 in the first UE group can decode their received RAR correctly, i.e., the RA procedure is likely to fail for UE3. Exhaustive numerical simulation however reveals that even with sufficiently large number of RA requests in a single RA slot (e.g. 30 RA requests/10 ms frame), with 17 RA preambles, the average fraction of UEs in a UE group, which do not contribute to the group common CIR is less than $10 \%$.}

\vspace{-0.6 cm}

\subsection{Automatic Contention Resolution}
\vspace{-0.3 cm}

Once the RAR is received at the UE on the designated subcarriers, the UE performs RAR decoding. Using this decoded RAR information, the UE then performs UL timing correction based on the received TA estimate. Next, using the scheduling information received in the RAR, the UE prepares for UL pilot and data transmission. Note that the RAR is usually CRC (cyclic redundancy check) protected. If the CRC check fails, the UE simply takes it as a RAR decoding failure. In such cases, the UE declares the current RA attempt to be unsuccessful and prepares for re-initiating the RA procedure with a new randomly selected RA preamble in the next RA uplink slot. {Note that the \textit{contention for resources amongst users is resolved automatically}, as the UEs, for which the RAR detection fails, cannot know the allocated uplink resource and therefore they would naturally back off from uplink data transmission.{\footnote[16]{{Note that in our proposed RA procedure, the step of RAR beamforming after user grouping and TA estimation is mandatory. This is because the strategy of not transmitting RAR when the probability of contention is high (in high UE density scenarios) would only increase the average RA latency, due to re-transmission of RA preambles by the users.}}} Also, it is possible that multiple UEs from the same UE group might be able to decode the RAR information block successfully. In such cases, the resources granted by the BS would be shared by all such UEs. To be precise, the large antenna array at the MaMi BS would allow for all such UEs to communicate simultaneously on the same uplink time-frequency resource.\footnote[17]{{With several tens of antennas at the MaMi BS, the channel rank is expected to be sufficiently high so that the BS would be able to separate the uplink messages from different UEs in the same UE group. To enable this, the UEs can use their unique core network identifier to choose mutually orthogonal pilots for transmission on the same shared UL resource. Previous study of the detection performance of such multi-user transmissions in MaMi uplink in \cite{Ngo2} reveals that a sufficiently large antenna array at the BS would help in separating out the signals received from different UEs.}}}

\vspace{-0.6 cm}

\subsection{SINR Analysis}\label{sinranalysis}
\vspace{-0.3 cm}

{In this section, our goal is to analyze the dependence of the received SINR on the number of UEs in a UE group as well as on the number of BS antennas. As the RAR corresponding to different RA preambles is transmitted on different orthogonal subcarriers, it suffices to consider the SINR analysis of the RAR transmission for the $k^{\text{th}}$ RA preamble only. We consider a worst case scenario, where the round-trip propagation delay is the same for all UEs detected on the $k^{\text{th}}$ RA preamble, i.e., their channel impulse response completely overlap in the time domain and also that there is only one UE group (i.e. $D_k = 1$). To focus only on the impact of multiple UEs on the received SINR at each UE, we consider perfect estimation of the group common TA, i.e., $\widehat{\tau}_{g,1} = \tau_{g,i}$, where $i = 1, 2, \ldots, K_g$ (note that $g = 1$ for the worst case scenario considered here).} {Substituting $\widehat{\tau}_{g,1} = \tau_{g,i}, \, \forall i = 1, 2, \ldots, K_g$ in \eqref{eq:fdchanest}, the group common CIR estimate is given by $\widetilde{H}_{m,g}[n] = \sqrt{\pu}\sum_{q=1}^{K_g}H_{mgq}[n] \, +  \, \frac{1}{\sqrt{\N}}W_{m,g}[n]$, where $W_{m,g}[n] \Define \frac{1}{\sqrt{N_{\text{RS}}}}\sum_{l=0}^{L-1}w_m[\widehat{\tau}_{g,1} + l + \xi_k]e^{-j\frac{2\pi}{N_{\text{RS}}}nl} \sim \mathcal{C}\mathcal{N}(0, \frac{L}{N_{\text{RS}}}\sigma^2)$. Using this expression of $\widetilde{H}_{m,g}[n]$ in \eqref{eq:rxdcig}, the received signal at the $i^{\text{th}}$ UE on the $n^{\text{th}}$ subcarrier is given by $Y_{g,i}[n] = \sqrt{\frac{N_{\text{RS}} P_{\text{d}}}{\upsilon_g}}\bm H_{gi}^T[n]\widetilde{\bm H}_g^{\ast}[n] u_g[n] \, + \, E_{g,i}[n]$, where $\upsilon_g = \Eb{||\widetilde{\bm H}_g[n]||^2} = M \Big (\pu \sum\limits_{q=1}^{K_g}\alpha_{gq} + \frac{L}{\N N_{\text{RS}}}\sigma^2\Big )$ and $\alpha_{gq} \Define \frac{1}{N_{\text{RS}}}\sum_{l=0}^{L-1}\sigma_{hgql}^2$. Using the expression of $\widetilde{\bm H}_g[n]$ in the expression of $Y_{g,i}[n]$ we get}

\vspace{-0.8 cm}

{\begin{IEEEeqnarray}{rCl}
\label{eq:toyrxsig}
\nonumber Y_{g,i}[n] & = & \sqrt{\frac{N_{\text{RS}} P_{\text{d}}}{\upsilon_g}} \,\,\, \bm H_{gi}^T[n]\Big(\sqrt{\pu}\sum_{q=1}^{K_g}\bm H_{gq}[n] \, +  \, \frac{1}{\sqrt{\N}}\bm W_{g}[n]\Big)^{\ast} u_g[n] \, + \, E_{g,i}[n]\\
\nonumber & = & \underbrace{\sqrt{\frac{N_{\text{RS}} P_{\text{d}}\pu}{\upsilon_g}} \,\, {||\bm H_{gi}[n]||^2}u_g[n]}_{\text{signal term}}\\
& & \, + \, \underbrace{\sqrt{\frac{N_{\text{RS}} P_{\text{d}}\pu}{\upsilon_g}}\bm H_{gi}^T[n]\sum\limits_{q=1, q \neq i}^{K_g}\bm H_{gq}^{\ast}[n] \, u_g[n] + \, \sqrt{\frac{N_{\text{RS}} P_{\text{d}}}{\upsilon_g \, \N}} \bm H_{gi}^T[n]\, \bm W_g^{\ast}[n] \, u_g[n] \, + \, E_{g,i}[n]}_{\Define \, \text{IN}_{g,i}[n] \, \, (\text{noise and interference})} \, ,
\IEEEeqnarraynumspace
\end{IEEEeqnarray}}

\vspace{-0.8 cm}

\noindent {where $P_d = \frac{N_{\text{RS}}}{N_{\text{SC}}}P_T$ (here $P_T$ is the total downlink transmit power for beamforming RAR to all UEs detected on the $k^{\text{th}}$ RA preamble). Here $\bm W_g[n] \Define (W_{1,g}[n], W_{2,g}[n], \cdots, W_{M,g}[n])^T$ and the last three terms on the R.H.S. of the second line of \eqref{eq:toyrxsig} are due to multi-user interference (MUI), channel estimation error and AWGN noise. Although, an expression for the instantaneous SINR in terms of the channel gains and the channel estimation noise can be derived from \eqref{eq:toyrxsig}, it turns out that this SINR expression is difficult to analyze due to which we cannot obtain insights about the variation of RA failure probability and RA latency. Therefore, we derive the long-term average SINR, which depends only on the statistics of the channel and noise and does not depend on any particular realization of the channel and noise. We therefore use the approach in \cite{Hasibi2,Hong6} to calculate the long-term average SINR. In this approach, in \eqref{eq:toyrxsig} we add and subtract the mean value of the signal term (i.e., $\text{DS}_{g,i}[n] \Define \, {\sqrt{\frac{N_{\text{RS}} P_{\text{d}}\pu}{\upsilon_g}} \,\, \Eb{||\bm H_{gi}[n]||^2}u_g[n]}$) to the RHS of \eqref{eq:toyrxsig}. The mean value becomes the new signal term and the variation around the mean (i.e., ${\sqrt{\frac{N_{\text{RS}} P_{\text{d}}\pu}{\upsilon_g}} \,\, (||\bm H_{gi}[n]||^2 - \Eb{||\bm H_{gi}[n]||^2})u_g[n]}$) is relegated to the other noise terms, i.e.,}

\vspace{-0.9 cm}

{\begin{IEEEeqnarray}{rCl}
\label{eq:toyrxsig2}
Y_{g,i}[n] & = &  \text{DS}_{g,i}[n]  \, + \, \underbrace{{\sqrt{\frac{N_{\text{RS}} P_{\text{d}}\pu}{\upsilon_g}} \,\, (||\bm H_{gi}[n]||^2 - \Eb{||\bm H_{gi}[n]||^2})u_g[n]} \, + \,  \text{IN}_{g,i}[n]}_{\Define \, \text{EN}_{g,i}[n]} \, .
\IEEEeqnarraynumspace
\end{IEEEeqnarray}}

\vspace{-0.75 cm}

\indent {We note that the signal term $\text{DS}_{g,i}[n]$ and the noise term $\text{EN}_{g,i}[n]$ in \eqref{eq:toyrxsig2} are uncorrelated and therefore the worst case scenario (in terms of information rate) is when the effective noise is Gaussian distributed, for which the information rate to the $i^{\text{th}}$ UE in the $g^{\text{th}}$ UE group is given by $\log_2(1 +\text{SINR}_{g,i}[n])$, where $\text{SINR}_{g,i}[n] \Define \frac{\Eb{|\text{DS}_{g,i}[n]|^2}}{\Eb{|\text{EN}_{g,i}[n]|^2}}$ is the long-term average SINR, i.e.,}

\vspace{-0.9 cm}

{\begin{IEEEeqnarray}{rCl}
\label{eq:toysinr}
{\text{SINR}_{gi}[n]} & = &  \left[{\frac{1}{M} \Big(1 + \frac{1}{N_{\text{RS}}\alpha_{gi} \gamma_d}\Big)\sum\limits_{q=1}^{K_g}\frac{\alpha_{gq}}{\alpha_{gi}} \, + \, \frac{L}{M \gamma N_{\text{RS}} \N \alpha_{gi}} \, + \, \frac{L}{M \gamma \gamma_d N_{\text{RS}}^2 \N \alpha_{gi}^2} }\right]^{-1}
\IEEEeqnarraynumspace
\end{IEEEeqnarray}}

\vspace{-0.9 cm}

\noindent {where $\gamma = \frac{\pu}{\sigma^2}$, $\gamma_{d} \Define \frac{N_{\text{RS}}}{N_{\text{SC}}}\frac{P_T}{\sigma^2}$ and $\alpha_{gq} = \frac{1}{N_{\text{RS}}}\sum_{l=0}^{L-1}\sigma_{hgql}^2$. In the following proposition, we derive the important result that in order to achieve a fixed target long-term average received SINR, both $\pu$ and $P_T$ can be decreased with increasing number of BS antennas, $M$.}

\vspace{-0.5 cm}

{\begin{proposition}
\label{PtvsM}
\normalfont For any given fixed desired long-term average value of the received SINR and fixed $K_g$, with $\pu \propto \frac{1}{\sqrt{M}}$, i.e., $\lim\limits_{M \to \infty}\sqrt{M}\pu = $ constant ($>0$), the required $P_T$ can also be decreased as $\frac{1}{\sqrt{M}}$ as $M \to \infty$, i.e., $\lim\limits_{M\to \infty}\sqrt{M}P_T = $ constant ($>0$).
\end{proposition}}

\vspace{-0.3 cm}

{\begin{IEEEproof}
{See Appendix~\ref{proofprop2}.\hfill \IEEEQEDhere}
\end{IEEEproof}}

\vspace{-0.5 cm}

\begin{figure}[t]
\vspace{-0.7 cm}
\hspace{-0.25 in} 
\subfloat[][]{\includegraphics[width= 3.4 in, height= 1.8 in]{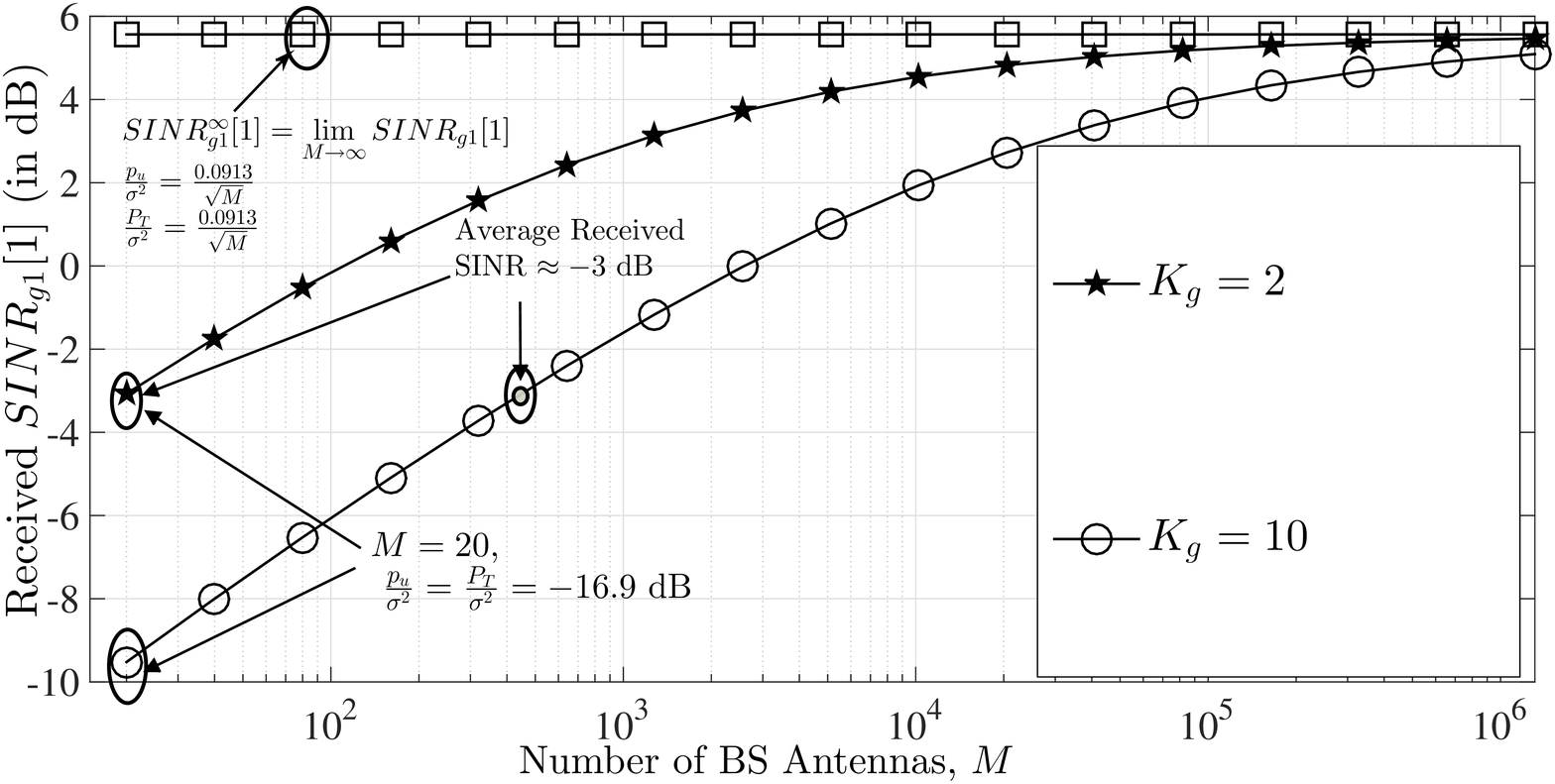}\label{fig:varMsinr}}
\subfloat[][]{\includegraphics[width= 3.4 in, height= 1.8 in]{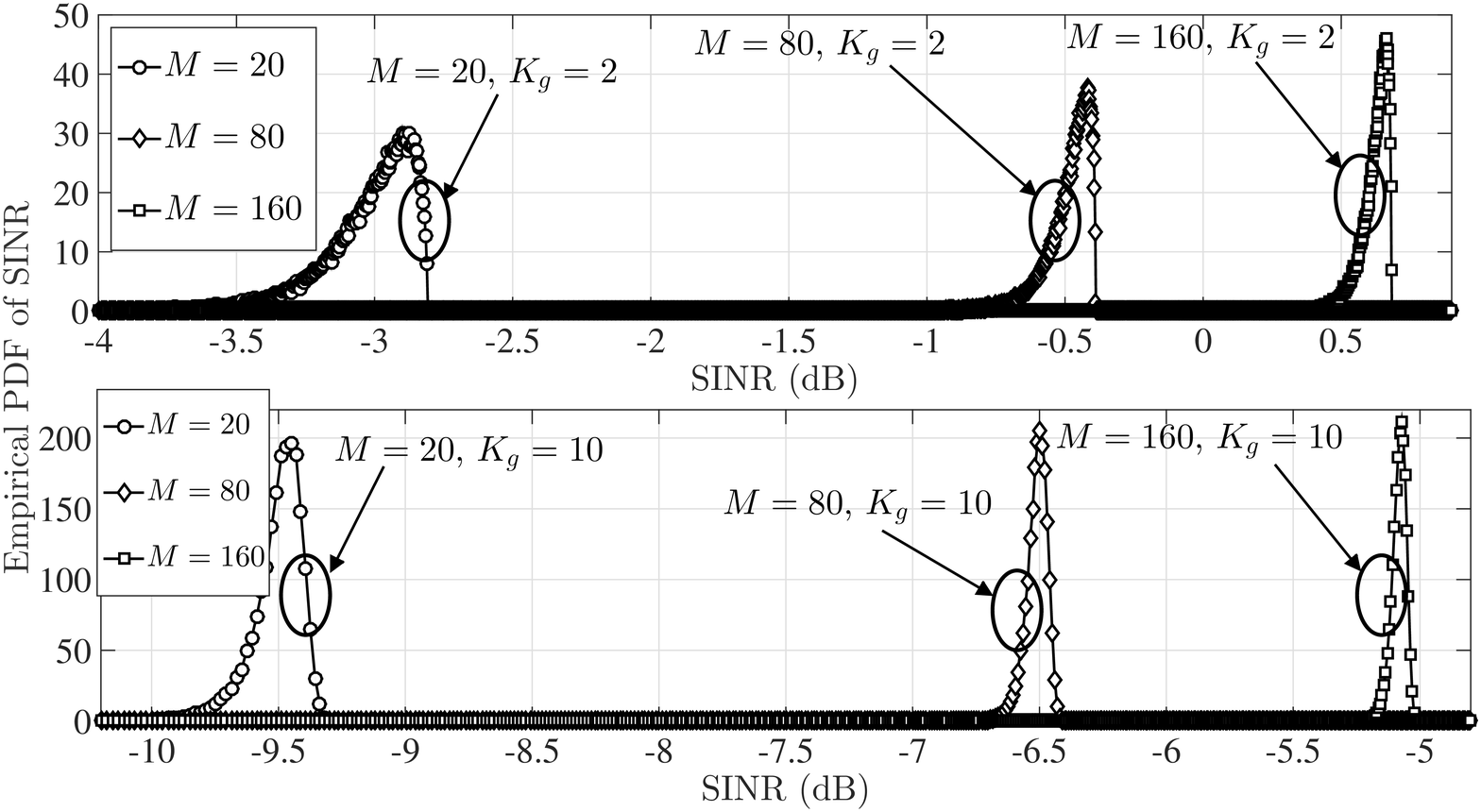}\label{fig:pdfsinr}}
\caption {{Plot of (a) Variation of ${\text{SINR}_{g1}[1]}$ as a function of increasing $M$, with $P_T \propto \frac{1}{\sqrt{M}}$, $\pu \propto \frac{1}{\sqrt{M}}$; (b) the Empirical pdf of the instantaneous received SINR for both low ($K_g = 2$) and high ($K_g = 10$) UE density scenarios ($L = 6$, $\N = 864$, $N_{\text{RS}} = 72$ and $N_{\text{SC}} = 24$).}}
\label{fig:toysinr}
\vspace{-0.2 cm}
\end{figure}

\begin{remark}
\label{sinrMrepeat}
\normalfont {This important result in Proposition~\ref{PtvsM} shows that for a fixed desired value of the received SINR, both the per-user RA preamble transmit power in the uplink and the total RAR beamforming power in the downlink can be decreased roughly by 1.5 dB with every doubling in the number of BS antennas $M$. This is supported in Fig.~\ref{fig:toysinr}~\subref{fig:varMsinr}, where we plot the variation in ${\text{SINR}_{g1}[n]}$ as a function of increasing $M$, with both $\pu$ and $P_T$ decreasing as $\frac{1}{\sqrt{M}}$.} {Substituting $\pu = \frac{\sigma^2 E_u}{\sqrt{M}}$ and $P_T = \frac{\sigma^2 E_T}{\sqrt{M}}$ in \eqref{eq:toysinr} we have}

\vspace{-1 cm}

{\begin{IEEEeqnarray}{rCl}
\label{eq:toysinrrev3}
{\text{SINR}_{gi}[n]} & = & \left[{\frac{1}{M} \Big(1 + \frac{N_{\text{SC}} \sqrt{M}}{N_{\text{RS}}^2\alpha_{gi} E_T}\Big)\sum\limits_{q=1}^{K_g}\frac{\alpha_{gq}}{\alpha_{gi}} \, + \, \frac{L}{\sqrt{M} E_u N_{\text{RS}} \N \alpha_{gi}} \, + \, \frac{L N_{\text{SC}}}{E_u E_T N_{\text{RS}}^3 \N \alpha_{gi}^2} }\right]^{-1} \, .
\IEEEeqnarraynumspace
\end{IEEEeqnarray}}

\vspace{-0.85 cm}

\noindent {Note that in the finite $M$ regime, with fixed $K_g$ and both $\pu$ and $P_T$ decreasing as $\frac{1}{\sqrt{M}}$, the first two terms (i.e. all terms except the last one) in the R.H.S. of \eqref{eq:toysinrrev3} decrease significantly with increasing $M$. Consequently the overall received SINR is observed to increase with increasing $M$. For example in Fig.~\ref{fig:toysinr}~\subref{fig:varMsinr}, with $K_g = 2$, ${\text{SINR}_{g1}[1]}$ increases roughly by $1.33$ dB as $M$ increases from $M = 20$ to $M = 40$.} {This increase in the average SINR can also be observed in Fig.~\ref{fig:toysinr}~\subref{fig:pdfsinr}, where we plot the empirical pdf of the instantaneous received SINR, with $\frac{\pu}{\sigma^2} = \frac{0.0913}{\sqrt{M}}$ and $\frac{P_T}{\sigma^2} = \frac{0.0913}{\sqrt{M}}$ (here $\pu$ is chosen so that for $M = 20$, the probability of TA estimation error in the contention-free scenario is $10^{-2}$). Note that due to channel hardening with increasing $M$, the variation of the empirical pdf around the mean also decreases. In other words, for any fixed UE density, the probability of having a very small SINR (relative to its mean value) at any UE decreases with increasing $M$ (see Fig.~\ref{fig:toysinr}~\subref{fig:pdfsinr}). Since successful RAR decoding (i.e. successful RA attempt) depends on the received SINR, the number of repeat RA attempts would therefore decrease with increasing number of BS antennas $M$, for a fixed UE density.  \hfill \qed}
\end{remark} 

\vspace{-0.4 cm}

{Using limit $M \to \infty$ on both sides of \eqref{eq:toysinrrev3} we have $\lim\limits_{M \to \infty} {\text{SINR}_{gi}[n]} = \frac{N_{\text{RS}}^3 \N E_u E_T \alpha_{gi}^2}{L N_{\text{SC}}} \Define \gamma_u$, i.e., when $M$ is sufficiently large, the average received SINR converges to a constant value which does not depend on $K_g$. In other words, as long as the desired received SINR is less than this asymptotic limit $\gamma_u$, it can be achieved by choosing an appropriate number of BS antennas $M$ for any value of $K_g$. In the following proposition, we compute this required value of $M$ for a given $K_g$ and show that it increases with increasing $K_g$ (i.e. equivalently UE density).}

\vspace{-.5 cm}

{\begin{proposition}
\label{finiteMvsKg}
\normalfont Let $\lim\limits_{M \to\infty}\sqrt{M}\frac{\pu}{\sigma^2} = c_1 > 0$, $\lim\limits_{M \to \infty}\sqrt{M}\frac{P_T}{\sigma^2} = c_2 > 0$ and system and channel parameters (i.e. $\N$, $N_{\text{SC}}$, $N_{\text{RS}}$ and $L$) be fixed. To achieve a fixed desired target ${\text{SINR}_{gi}[n]} = \epsilon < \gamma_u$, the minimum required number of BS antennas denoted by $M^{\star}(K_g, c_1, c_2, \epsilon)$ would increase with increasing number of UEs, $K_g$ (here $\gamma_u = \lim\limits_{M \to\infty}{\text{SINR}_{gi}[n]} = \frac{N_{\text{RS}}^3 \N c_1 c_2 \alpha_{gi}^2}{L N_{\text{SC}}}$).
\end{proposition}}

\vspace{-0.5 cm}

{\begin{IEEEproof}
See Appendix~\ref{finiteMvsKgproof}. \hfill \IEEEQEDhere
\end{IEEEproof}}

\vspace{-0.3 cm}

\par {In the following we intuitively explain the above result in Proposition~\ref{finiteMvsKg}. In the finite $M$ regime, the first term in the denominator of \eqref{eq:toysinrrev3} increases as the number of UEs, $K_g$ increases and therefore for a fixed $M$, the effective average received SINR decreases with increasing $K_g$. Since this term decreases as $\frac{1}{\sqrt{M}}$ with increasing $M$, we can compensate for the reduction in SINR (due to increasing $K_g$) by increasing $M$ to a sufficiently large value. This phenomenon is also observed in Fig.~\ref{fig:toysinr}~\subref{fig:varMsinr}, {where for the same fixed desired average SINR of -3 dB, the number of BS antennas required for $K_g = 2$ (low UE density scenario) is only $M = 20$, while for $K_g = 10$ (high UE density scenario), it is $M \approx 410$. This shows the robustness of our proposed RA method at high user densities, as it can achieve any  fixed target received SINR less than $\gamma_u$  by increasing the size of the antenna array at the BS.}}

\vspace{-0.7 cm}

\section{Numerical Analysis}

\vspace{-0.5 cm}

In this section, we use Monte-Carlo simulations to study the performance of the above proposed RA procedure (i.e. TA estimation and RAR beamforming) for random access in TDD MaMi systems. For our simulation, we assume that the total PRACH bandwidth is $1.08$ MHz and the subcarrier spacing in the physical uplink shared channel (PUSCH) is $15$ KHz, while the PRACH subcarrier spacing for uplink RA transmission is $1.25$ KHz (same as in LTE systems \cite{Skold}).{\footnote[18]{{This 1.25 KHz subcarrier spacing in PRACH ensures that the ZC sequence used for RA preamble design is of duration $\frac{1}{1.25 \text{KHz}} = 0.8$ ms, so that it fits within 1 ms LTE subframe along with the guard time, which is usually equal to the maximum round-trip delay of the cell. Also, keeping the shared channel subcarrier spacing to be an integer multiple of the PRACH subcarrier spacing minimizes the orthogonality loss between the PRACH and the PUSCH resources \cite{Baker}}.}} Thus, the total number of PRACH subcarriers for UL transmission is $\big \lfloor \frac{1.08 \text{MHz}}{1.25 \text{KHz}} \big \rfloor = 864$. We also assume that $\N = 864$ and the cell radius is $6$ km. Therefore the maximum round-trip propagation delay for the cell would be $6 \times 6.7 = 40.2 \mu$s (with $6.7 \mu$s per km round-trip propagation delay \cite{Skold}). Assuming the maximum delay spread of the wireless channel to  be $5 \mu$s, we have $L = \left\lceil 1.08 \times 5 \right\rceil = 6$ channel uses. Hence, the length of the cyclic prefix (CP) for uplink RA preamble transmission would be $G = \lceil 1.08 (40.2 + 5) \rceil = 50$ channel uses and the total number of distinct RA preambles therefore would be $\big \lfloor \frac{0.8 \, \text{ms}}{(40.2 + 5) \mu\text{s}}\big \rfloor = 17$.

\vspace{-0.8 cm}

\subsection{Density and Distribution of User Location}

\vspace{-0.3 cm}

For simulation purposes, we model the locations of UEs requesting random access as a homogeneous Poisson Point Process (PPP) with the cellular BS at the origin. Note that our proposed user grouping, TA estimation and RAR beamforming procedure is for random access in \emph{crowded} massive MIMO scenario. For instance with $2.26 \times 10^6$ devices in a $6$ km radius cell (i.e. device density of 20000/sq.km), if each device makes a RA attempt every $12$ minutes on an average, then the average number of RA requests in a duration of $10$ ms (assuming $1$ RA slot in a $10$ ms frame) would be $\approx 31.42$.

\vspace{-0.8 cm}

\subsection{Design \& Transmission of Random Access Response}

\vspace{-0.3 cm}

{We assume that the random access response (RAR) for any UE group contains the following set of information: (a) the RA acknowledgement bit (a logical \lq{1}' bit repeated 7 times); (b) the TA information (for $40.2 \, \mu$s maximum delay spread and $1.08$ MHz PRACH, the maximum value of TA is $44$ channel uses which is represented using $6$ bits); (c) for the UL resource allocation, the starting resource block (RB) index is transmitted and assuming 2.7 MHz of uplink channel, we have 15 RBs.\footnote[19]{{In LTE a resource block contains 12 shared channel subcarriers, i.e., it has a bandwidth of 180 KHz \cite{Skold}.}} Clearly, the starting RB index would require 4 bits; and (d) two bits to represent the number of RBs allocated for subsequent transmissions (assuming the BS allocates at most 4 RBs). Note that the actual information in the RAR is contained within these $6 + 4 + 2 = 12$ bits, which is then subsequently CRC coded with CCITT-5 CRC polynomial \cite{CCITT}. The 7 RA acknowledgement bits are then appended at the beginning of this CRC coded sequence, thus forming a 24 bit random access response (RAR). This 24 bit RAR is then BPSK modulated and beamformed as discussed in Section~\ref{rarbeam}. As the same subcarriers in PRACH are used for RAR beamforming, we have $N_{\text{RS}} = \big \lfloor \frac{1.08 \, \text{MHz}}{15 \, \text{KHz}}\big \rfloor = 72$ shared channel subcarriers and a 1 ms subframe for RAR transmission (i.e. 14 OFDM symbols as in LTE). Therefore the total number of time-frequency resource elements (REs) available for RAR transmission is $72 \times 14 = 1008$, whereas in the worst case (with RA request on each RA preamble) the number of RAR bits required to be transmitted for all 17 RA preambles is only $17 \times N_{\text{SC}} = 408$, since each bit of the 24-bit RAR is transmitted on a different subcarrier. Clearly, as the total number of required resource elements (i.e. 408 REs) is much smaller than the number of available REs (i.e. 1008), we can use frequency hopping patterns to repetitively transmit the RAR sequences in order to exploit frequency diversity. For RAR detection and decoding, any UE requesting RA, would first attempt to detect the RA acknowledgement bits. If the number of logical \lq{1}'s detected is more than 4, then the UE assumes that it has received a RAR. Upon detection of RAR, the UE would check the CRC. If the CRC check fails, the UE marks the RAR decoding attempt as unsuccessful. Otherwise if the CRC check is validated, the UE assumes successful RAR decoding and uses the decoded RAR for UL timing correction and subsequent user identity transmission on the allocated uplink resource mentioned in the RAR. Therefore in our proposed RA procedure, a UE would declare its RA attempt to be unsuccessful if it does not detect any RAR or if the RAR decoding fails. After an unsuccessful RA attempt, the UE would re-initiate RA with a new randomly selected RA preamble in the next available RA UL slot.}

\vspace{-0.75 cm}

\subsection{Results \& Discussions}
\label{results}
\vspace{-0.3 cm}

\begin{figure}[t]
\vspace{-0.45 in} 
\subfloat[][]{\includegraphics[width= 3.2 in, height= 1.9 in]{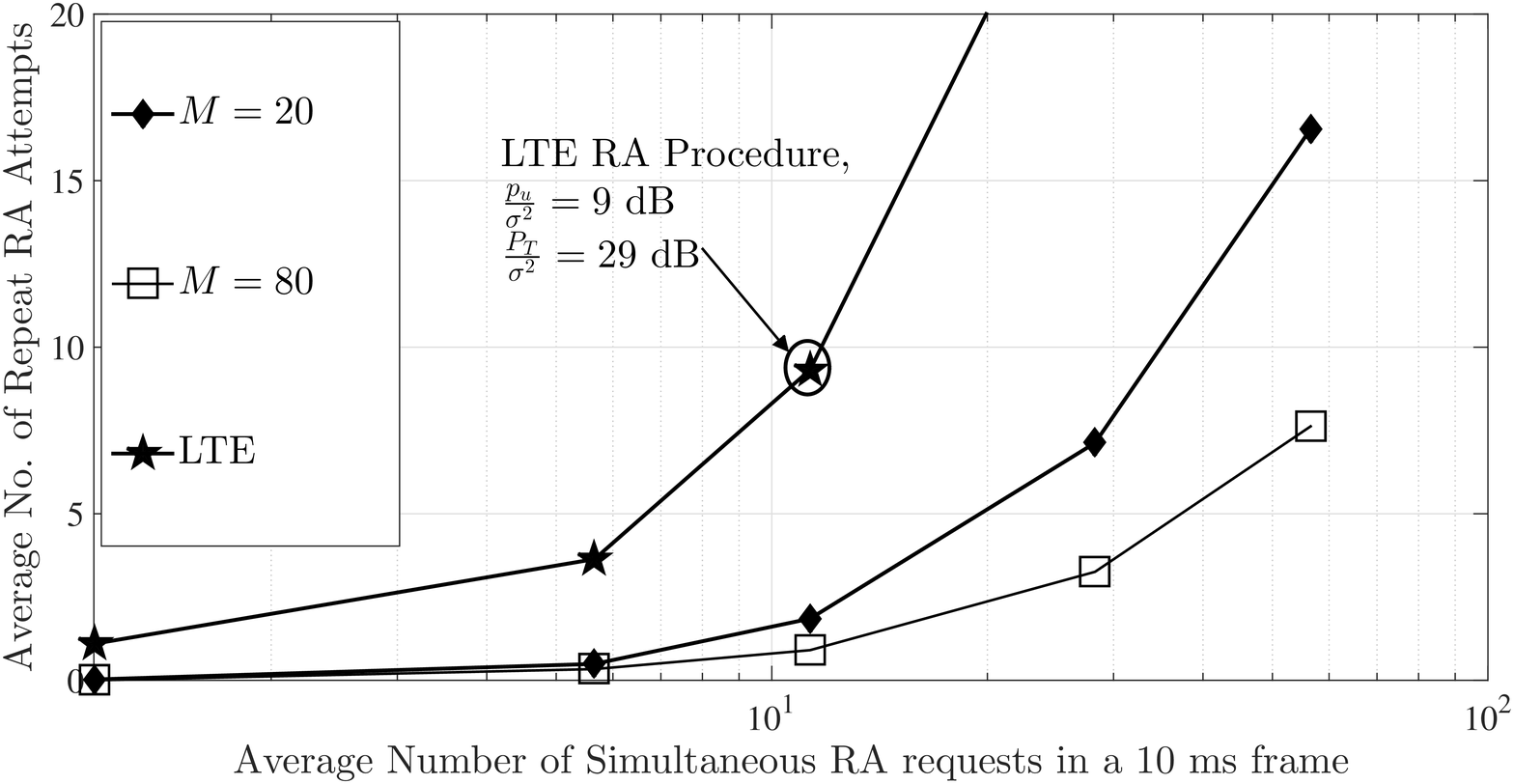}\label{fig:navgvsue}}
\subfloat[][]{\includegraphics[width= 3.2 in, height= 1.9 in]{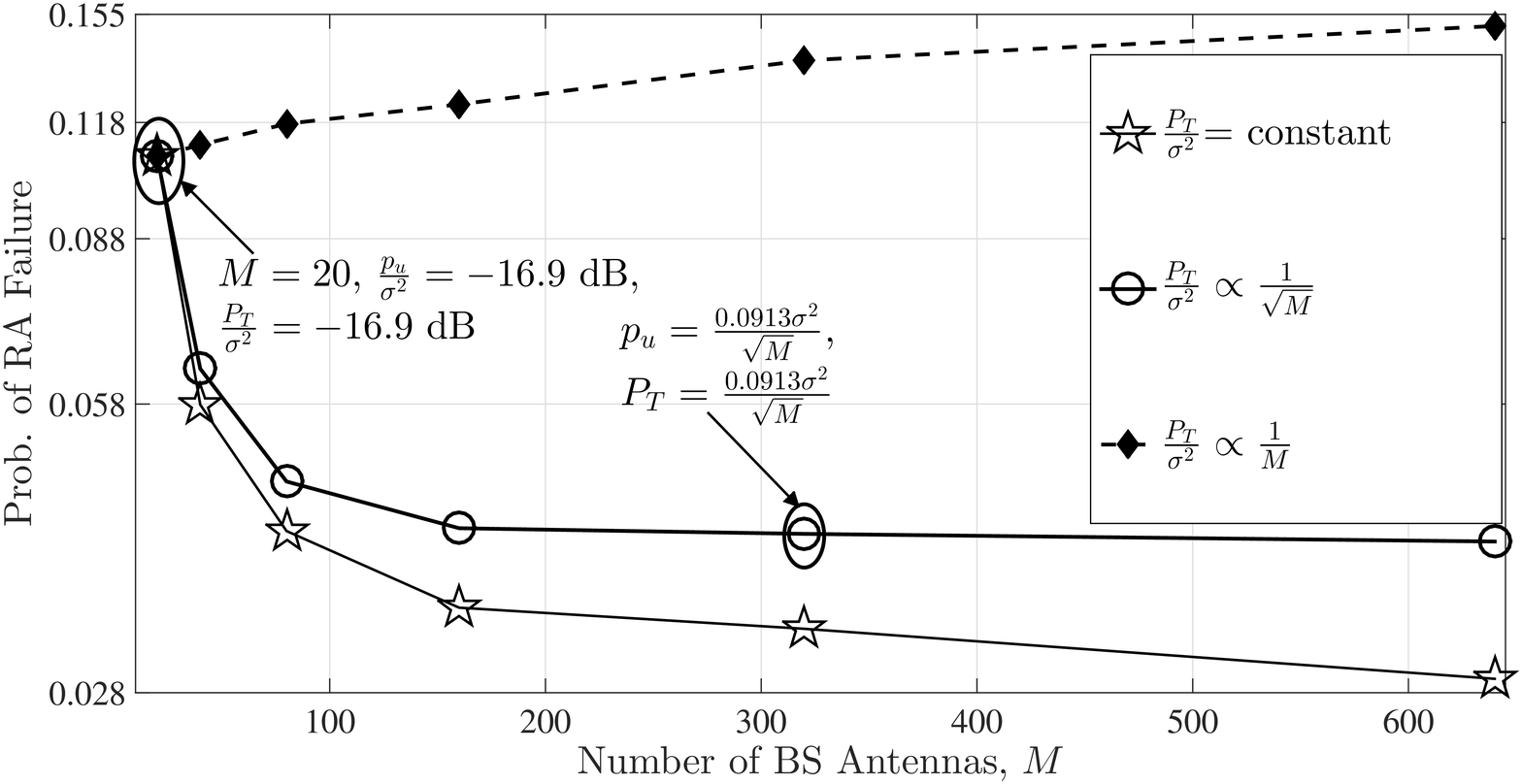}\label{fig:RAfailprob}}
\caption {{(a) Plot of the average number of repeat RA attempts as a function of increasing number of simultaneous RA requests, for $M = 20$ and $80$; (b) Plot of the prob. of RA failure (number of repeat RA attempts $> \, 5$) versus the number of BS antennas, $M$, for a fixed average number of simultaneous RA requests = 11 in a 10 ms frame. }}
\label{fig:numsec}
\vspace{-0.6 cm}
\end{figure}

Using the above RAR design and the proposed RA procedure, in this section we study the following: (a) the impact of increasing UE density and also increasing number of BS antennas on the average number of repeat RA attempts; and (b) the impact of increasing number of BS antennas on the probability of RA failure for a fixed UE density. To study the impact of increasing UE density on the average number of repeat RA attempts, in Fig.~\ref{fig:numsec}~\subref{fig:navgvsue}, we plot the average number of repeat RA attempts as a function of increasing number of simultaneous RA requests in a 10 ms frame, for $M = 20$ and $80$ BS antennas. For this simulation, we assume that both the per-user RA preamble transmit power $\pu$ and the total downlink beamforming power $P_T$ are fixed (e.g., $\frac{\pu}{\sigma^2} = \frac{P_T}{\sigma^2} = -16.9$ dB) with increasing number of BS antennas $M$. It is observed that for any given $M$, the average number of repeat RA attempts\footnote[20]{Note that the number of repeat RA attempts is equal to the number of extra attempts (not counting the first attempt) made by the UE, till it is able to successfully decode the RAR.} increases with increasing number of simultaneous RA requests (i.e. equivalently increasing UE density). This is expected since with increasing number of simultaneous RA requests, the number of UEs in any UE group is expected to increase and therefore for a fixed $M$, there would be more MUI in the received RAR (see the discussion in the paragraph following Proposition~\ref{finiteMvsKg}). {In Fig.~\ref{fig:numsec}~\subref{fig:navgvsue}, we also plot the number of repeat RA attempts required for the LTE RA procedure, which is not only observed to be significantly larger compared to that of our proposed RA procedure but also requires much higher RA preamble and RAR transmit power. For instance, with an average of 11 RA requests per 10 ms frame, the average number of repeat RA attempts required for LTE is $\approx 8.3$, while for our proposed RA procedure (with $M = 20$ BS antennas), it is only $1.8$. Further, from Fig.~\ref{fig:numsec}~\subref{fig:navgvsue}, it is also observed that for a fixed number of repeat RA attempts, a MaMi BS with a larger number of BS antennas can successfully handle a much larger number of RA requests. For instance, for a fixed average number of repeat RA attempts equal to $1.8$, the average number of simultaneous RA requests that can be handled is $\approx 11$ with $M = 20$ BS antennas, while with $M = 80$ BS antennas, a larger number of RA requests ($\approx 16.5$) can be handled. This is due to the fact that a larger number of BS antennas compensates for the extra MUI introduced when the number of RA requests increases. This therefore demonstrates the robustness of our proposed TA estimation, user grouping and RAR beamforming method for handling a large number of RA requests in crowded MaMi systems.}

\par RA failure for a UE happens when the UE is unable to successfully complete its RA procedure even after 5 repeat RA attempts. In Fig.~\ref{fig:numsec}~\subref{fig:RAfailprob}, we plot the numerically computed probability of RA failure as a function of increasing $M$, for a fixed average number of simultaneous RA requests ($\approx 11$) in a 10 ms frame. With $\frac{\pu}{\sigma^2} = \frac{0.0913}{\sqrt{M}}$, we plot the RA failure probability for the following three scenarios: (a) the total downlink transmit power $P_T$ decreases as $\frac{1}{M}$ (the curve with filled diamonds); (b) $P_T$ decreases as $\frac{1}{\sqrt{M}}$ (the curve with circles); and (c) $P_T$ remains constant (the curve with stars). It is observed that with constant $P_T$, the probability of RA failure decreases as $M$ increases. This is due to the increase in the average received SINR with increasing $M$. However, when the total downlink transmit power for RAR beamforming is reduced as $\frac{1}{\sqrt{M}}$, from the figure we observe that the RA failure probability converges to a non-zero constant. This observation is supported by Proposition~\ref{PtvsM}, where we know that if $P_T \propto \frac{1}{\sqrt{M}}$ as $M \to \infty$, then the average received SINR converges to a non-zero constant value. On the other hand, with $P_T$ decreasing at a rate faster than $\frac{1}{\sqrt{M}}$ (e.g. when $P_T \propto \frac{1}{M}$ in Fig.~\ref{fig:numsec}~\subref{fig:RAfailprob}), it is observed that the probability of RA failure increases with increasing $M$. From these observations in Fig.~\ref{fig:numsec}~\subref{fig:RAfailprob} and the SINR analysis in Section~\ref{sinranalysis}, we conclude that for a fixed desired probability of RA failure, the minimum required $P_T$ and $\pu$ can both be decreased roughly by $1.5$ dB, with every doubling in the number of BS antennas. This is interesting since this is same as the best achievable power gain in TDD MaMi systems \cite{Ngo1}.


%
%

\vspace{-0.5 cm}

\appendix

\vspace{-0.5 cm}

\subsection{{Proof of Proposition~\ref{PfavsM}}}\label{falsealarm}

\vspace{-0.5 cm}

{In the absence of RA attempts using the $k^{\text{th}}$ RA preamble, from \eqref{eq:tdmod} and \eqref{eq:threshold} it follows that a false alarm event would occur, if and only if $P_k[t] > 0$ for some $t \in [0, G-1]$, i.e.,}
 
\vspace{-1.2 cm}

{\begin{IEEEeqnarray}{rCl}
\label{eq:falsealarm}
\nonumber P_F & \Define & \text{Pr}\{\omega_t > \theta_0 \,\, \text{for some}\, t \in [0, G-1]\}\, = \, 1 - \text{Pr}\{\omega_t \leq \theta_0, \, \forall t \in [0, G-1]\}\\
& = & 1 - [\text{Pr}\{\omega_t \leq \theta_0\}]^G\, = \, 1 - [1 - \text{Pr}\{\omega_t > \theta_0\}]^G \, ,
\IEEEeqnarraynumspace
\end{IEEEeqnarray}}

\vspace{-1.1 cm}

\noindent {since $\omega_t$ are all i.i.d., with mean = 0 and variance $\frac{\sigma^4}{M}$ (see \eqref{eq:varomegaeta}). Clearly we have}

\vspace{-0.85 cm}

{\begin{IEEEeqnarray}{rCl}
\label{eq:Pfalim1}
\Eb{\omega_t^2} & = & \int\limits_{-\infty}^{\infty}x^2 f_{\omega_t}(x) dx \, \geq \, \int\limits_{\theta_0}^{\infty}x^2 f_{\omega_t}(x) dx \, \geq \, \theta_0^2 \, \int\limits_{\theta_0}^{\infty} f_{\omega_t}(x) dx \, = \, \theta_0^2 \, \text{Pr}\{\omega_t > \theta_0\} \, . 
\IEEEeqnarraynumspace
\end{IEEEeqnarray}}

\vspace{-0.7 cm}

\indent {Here $f_{\omega_t}(x)$ is the pdf of the random variable $\omega_t$. In other words, from \eqref{eq:Pfalim1}, we have $\text{Pr}\{\omega_t > \theta_0\} \leq \frac{1}{\theta_0^2}\Eb{\omega_t^2} =  \frac{\sigma^4}{M \theta_0^2}$. Substituting this result in \eqref{eq:falsealarm}, we get}

\vspace{-1 cm}

{\begin{IEEEeqnarray}{rCl}
\label{eq:Pfalim2}
P_F & \leq & 1 - \Big[1 - \frac{\sigma^4}{M \theta_0^2} \Big]^G \, = \, 1 - \Big[1 - \frac{1}{\kappa^2}\Big]^G \, ,
\IEEEeqnarraynumspace
\end{IEEEeqnarray}}

\vspace{-0.9 cm}

\noindent {for $\theta_0 = \kappa \frac{\sigma^2}{\sqrt{M}}$ ($\kappa > 1$) (as given in the statement of the proposition).}

\vspace{-0.6 cm}




\subsection{{Proof of Proposition~\ref{PtvsM}}}\label{proofprop2}

\vspace{-0.4 cm}

{Substituting $\gamma = \frac{\pu}{\sigma^2} = \frac{E_u}{\sqrt{M}}$ in \eqref{eq:toysinr}, we have}

\vspace{-0.85 cm}

{\begin{IEEEeqnarray}{rCl}
\label{eq:toysinrrev}
{\text{SINR}_{gi}[n]} & = & \left[{\frac{1}{M} \Big(1 + \frac{1}{N_{\text{RS}}\alpha_{gi} \gamma_d}\Big)\sum\limits_{q=1}^{K_g}\frac{\alpha_{gq}}{\alpha_{gi}} \, + \, \frac{L}{\sqrt{M} E_u N_{\text{RS}} \N \alpha_{gi}} \, + \, \frac{L}{\sqrt{M} \gamma_d E_u N_{\text{RS}}^2 \N \alpha_{gi}^2} }\right]^{-1} \, .
\IEEEeqnarraynumspace
\end{IEEEeqnarray}}

\vspace{-0.8 cm}

\indent {Assuming the received average SINR to be fixed, i.e., ${\text{SINR}_{gi}[n]} = \epsilon > 0$, from \eqref{eq:toysinrrev}, we obtain the following expression for $\gamma_d$, i.e.,}

\vspace{-0.8 cm}



{\begin{IEEEeqnarray}{rCl}
\label{eq:toysinrrev2}
\gamma_d & = & \frac{\frac{1}{M N_{\text{RS}}}\sum_{q=1}^{K_g}\frac{\alpha_{gq}}{\alpha_{gi}^2}\, + \, \frac{L}{\sqrt{M} E_u N_{\text{RS}}^2 \N \alpha_{gi}^2}}{\frac{1}{\epsilon} \, - \, \frac{1}{M}\sum_{q=1}^{K_g}\frac{\alpha_{gq}}{\alpha_{gi}} \, - \, \frac{L}{\sqrt{M} E_u N_{\text{RS}} \N \alpha_{gi}} } \, .
\IEEEeqnarraynumspace
\end{IEEEeqnarray}}

\vspace{-0.7 cm}

\indent {Substituting $\gamma_d = \frac{N_{\text{RS}} P_T}{N_{\text{SC}} \, \sigma^2}$ and multiplying both sides of \eqref{eq:toysinrrev2} by $\sqrt{M}$ and taking limit as $M \to \infty$, we have $\lim\limits_{M \to \infty} \sqrt{M}P_T = \frac{L N_{\text{SC}} \, \epsilon \, \sigma^2}{E_u N_{\text{RS}}^3 \N \alpha_{gi}^2} \, (\text{constant})$.}

%
%


\subsection{{Proof of Proposition~\ref{finiteMvsKg}}}
\label{finiteMvsKgproof}

{Substituting $\pu = \frac{c_1 \sigma^2}{\sqrt{M}}$ and $P_T = \frac{c_2 \sigma^2}{\sqrt{M}}$ in \eqref{eq:toysinr} for a fixed $\text{SINR}_{gi}[n] = \epsilon < \gamma_u$, we have}

\vspace{-0.9 cm}

{\begin{IEEEeqnarray}{rCl}
\label{eq:finiteMKg}
\frac{1}{\epsilon} & = & \frac{a_1}{M} + \frac{a_2 + a_3}{\sqrt{M}} + \frac{1}{\gamma_u} ,
\end{IEEEeqnarray}}

\vspace{-0.85 cm}

\noindent {where $a_1 = \sum_{q=1}^{K_g}\frac{\alpha_{gq}}{\alpha_{gi}}$, $a_2 \Define \frac{N_{\text{SC}}}{c_2 \alpha_{gi}N_{\text{RS}}^2}\sum_{q=1}^{K_g}\frac{\alpha_{gq}}{\alpha_{gi}}$, $a_3 \Define \frac{L}{\N N_{\text{RS}} \alpha_{gi} c_1}$, and $\frac{1}{\gamma_u} = \frac{L N_{\text{SC}}}{c_1 c_2 \N N_{\text{RS}}^3 \alpha_{gi}^2}$. Equation \eqref{eq:finiteMKg} is quadratic in $\sqrt{M}$ and has a unique solution for $M$ which is given by}

\vspace{-0.8 cm}

{\begin{IEEEeqnarray}{rCl}
\label{eq:solnM}
M^{\star}(K_g, c_1, c_2, \epsilon) & = & \Bigg[\frac{ (a_2 + a_3) + \sqrt{ (a_2 + a_3)^2 + 4a_1(\frac{1}{\epsilon} - \frac{1}{\gamma_u})}}{2(\frac{1}{\epsilon} - \frac{1}{\gamma_u})}\Bigg]^2 \, .
\end{IEEEeqnarray}
}

\vspace{-0.8 cm}

\indent {From the definition of $a_1$ and $a_2$ above, it is clear that as $K_g$ increases, the terms $a_1$ and $a_2$ would also increase and therefore from \eqref{eq:solnM} it follows that the required $M$ would also increase.}

\vspace{-0.6 cm}

\ifCLASSOPTIONcaptionsoff
  \newpage
\fi



%


\bibliographystyle{IEEEtran}
\bibliography{IEEEabrvn,mybibn}


\end{document}